\documentclass[11pt]{article}
\usepackage{latexsym}
\usepackage{amsmath}
\usepackage{amsthm}
\usepackage{amsfonts}
\allowdisplaybreaks
\title{Mixing-Matrix Renormalization Revisited}
\author{Antonio O.\ Bouzas \thanks{E-mail:
    abouzas@mda.cinvestav.mx}\\\small Departamento de F\'{\i}sica
  Aplicada, CINVESTAV-IPN \\\small Carretera Antigua a Progreso Km.\
  6, Apdo.\ Postal 73 ``Cordemex''\\\small
  M\'erida 97310, Yucat\'an, M\'exico}
\date{}
\newcommand{\ms}{ \ensuremath{\overline{\mathrm{MS}}} }
\newcommand{\mat}[1]{\ensuremath{\boldsymbol{#1}}}
\newcommand{\LL}{\ensuremath{\mathcal{L}}}
\newcommand{\sss}[1]{{\scriptscriptstyle #1}}
\newcommand{\naught}{\sss{0}}
\newcommand{\dirac}{\not\negmedspace\partial}
\newcommand{\pirac}{\not\negmedspace p\,}
\newcommand{\lirac}{\not\negmedspace \ell\,}
\newcommand{\tr}{\mathrm{Tr}}
\newcommand{\os}[1]{#1^\mathrm{(os)}}
\newcommand{\mub}{\overline{\mu}}
\theoremstyle{remark}
\newtheorem*{lem*}{Lemma}
\newtheorem{lem}{Lemma}

\begin{document}

\maketitle
\begin{abstract}
  We study the renormalization of normal mixing matrices, which
  includes hermitian and unitary matrices as particular cases.  We
  give a minimal, multiplicative parametrization of counterterms, and
  compute the renormalized Lagrangian to one-loop order in several
  simple models with $N$ species of fermions, both in on-shell
  and \ms schemes.  In on-shell scheme the mass-degenerate case is
  considered separately. 
\end{abstract}
\section{Introduction}
\label{sec:intro}
In theories with many particle species which mix non-trivially due to
interactions, the mixing matrix generally requires renormalization
like any other parameters in the Lagrangian.  Such renormalization has
been considered in \cite{sac} within the context of the Standard Model
(SM), and in extensions of the SM with Majorana neutrinos in
\cite{kni}.

In this paper we consider mixing-matrix renormalization in a more
generic setting.  Specifically, we study the renormalization of normal
mixing matrices (i.e., matrices commuting with their adjoint), which
includes hermitian and unitary matrices as particular cases.  We give
a minimal, multiplicative parametrization of counterterms, and compute
the renormalized Lagrangian to one-loop order in several simple models
with $N$ species of fermions, both in on-shell (OS) and \ms schemes.
In on-shell scheme the mass-degenerate case is considered separately.
We work in dimensional regularization \cite{gia} throughout this
paper.

Mixing-matrix renormalization in the SM, and in its extensions and
low-energy effective theories, is usually closely related to other
issues such as renormalization of theories with unstable particles,
gauge invariance of the renormalization procedure, and CP violation.
The latter is out of the scope of this paper.  We assume all particles
to be stable.  If that were not the case, one-loop self-energy parts
should be replaced by their dispersive parts in OS renormalization
conditions \cite{den}.  Gauge invariance of the renormalized CKM
matrix in the SM has been considered in \cite{gam,bar}.  In our case,
Abelian gauge invariance plays a role in the discussion of unitary
mixing matrices in \S \ref{sec:unitary}.

In the next section we consider a model with $N$ fermion flavors
coupled to a scalar particle through a Yukawa interaction, specified
by a hermitian coupling matrix.  Renormalization of this
model, and in particular the structure of its counterterms, is
considered in detail and computed at the one-loop level.  The results
are generalized to normal matrices in \S \ref{sec:normal}.  The
particular case of unitary interaction matrices is treated in \S
\ref{sec:unitary}, where the relation between our parametrization of
counterterms and the one commonly in use in the literature is
discussed.  In \S \ref{sec:final} we give some final remarks.  We
gather material relevant to all sections in four appendices.  In
appendix \ref{sec:appb}, in particular, we give a parametrization for
mappings of normal matrices that we find useful in discussing
renormalization of mass and mixing matrices.

\section{Hermitian mixing matrix}
\label{sec:hermitean}

The simplest model of fermion mixing consists of $N$ Dirac fields
$\psi_a$, $a=1,\dots, N$, which we gather together in a column field
$\mat{\psi}$, interacting with a scalar field $\phi$ through a Yukawa
coupling.  The Lagrangian is given by,
\begin{equation}
  \label{lag}
  \LL = -\frac{1}{2} \phi_\naught (\Box+{m_\phi}_\naught^2) \phi_\naught +
  \overline{\mat{\psi}}_\naught (i \dirac - \mat{M}_\naught) 
  \mat{\psi}_\naught + \overline{\mat{\psi}}_\naught \mat{H}_\naught
  \mat{\psi}_\naught \phi_\naught 
  -\frac{\xi_\naught}{3!} \phi_\naught^3 - \frac{\lambda_\naught}{4!}
  \phi_\naught^4,  
\end{equation}
the subindex 0 indicating bare fields and parameters.  The Yukawa
interaction term is specified by a matrix of couplings $\mat{H}$,
whose elements are the expansion parameters in perturbation theory.
$\mat{M}$ is the fermion mass matrix, which is assumed to be regular
(i.e., no two of its eigenvalues are equal).  The degenerate case will
be considered below in \S \ref{sec:degenerate}.  If
$[\mat{M},\mat{H}]\neq 0$, the interaction mixes flavors.  Clearly,
\mat{M} and \mat{H} must be hermitian.  The interaction terms in
$\phi^3$ and $\phi^4$ are needed for renormalizability.  If fermions
are massless and $\xi=0$, $\LL$ is invariant under the discrete
symmetry $\mat{\psi}\rightarrow \gamma_5\mat{\psi}$, $\phi\rightarrow
-\phi$ which forbids a $\phi^3$ term.  For massive fermions such a
term cannot be avoided.  Because of its simplicity and of its lack of
symmetries preventing renormalization of its parameters, $\LL$ is in
some sense archetypal, so we consider its renormalization in detail
below.

We never use the summation convention for flavor indices $a, b, c,
\dots$.  Space-time indices are denoted by Greek letters, with
summation over repeated indices always understood.  

We write $\LL$ in terms of renormalized fields and couplings by
introducing renormalization constants.  The field $\phi$ renormalizes
multiplicatively $\phi_\naught = Z_\phi^{1/2}\phi = (1+1/2 \,\delta
Z_\phi) \phi$.  In \ms $Z_\phi=Z_\phi(\mat{H}, \lambda,
d)$\cite{col1}, whereas in OS $Z_\phi=Z_\phi(\mat{H}, \lambda, d;
m_\phi^2/\mu^2, \mat{M}/\mu, \xi/\mu)$.  We do not expect
multiplicative renormalization of $\lambda$, which already at one loop
receives ${\cal O} (\mat{H}^4)$ divergent contributions from a box
diagram, $\lambda_\naught = \mu^\epsilon (\lambda+\delta \lambda)$.
The dependence of $\delta\lambda$ on the parameters of the theory is
analogous to that of $Z_\phi$.  In these expressions $d=4-\epsilon$ is
the dimension of space-time and $\mu$ the mass scale of dimensional
regularization.

The mass $m_\phi^2$ will mix under renormalization with the other
dimensionful parameters in $\LL$.  By dimensional analysis,
\[  
{m_\phi}_\naught^2 = Z_{m_\phi}^{(1)} m_\phi^2 +
\sum_{abcd} \delta Z_{abcd}^{(2)} M_{ab} M_{cd} + \delta
Z_{m_\phi}^{(3)} \xi^2.
\]
In OS renormalized masses are physical, so $M_{ab}=
{m_a}_\mathrm{ph}  \delta_{ab}$ and,
\[  
{m_\phi}_\naught^2 = Z_{m_\phi}^{(1)} {m_\phi}_\mathrm{ph}^2 + 
\sum_{ab} \delta Z_{ab}^{(2)} {m_a}_\mathrm{ph} {m_b}_\mathrm{ph}
+ \delta Z_{m_\phi}^{(3)} \xi^2, 
\]
the renormalization constants $\delta Z_{m_\phi}^{(j)}$ depending on
$\mat{H}$, $\lambda$, $d$ and ratios of dimensionful parameters.  In
\ms there is no dependence of $\delta Z_{m_\phi}^{(j)}$ on masses and
$\xi$ \cite{col1}, but $\mat{M}$ is not diagonal.  We can choose our
flavor basis, however, so that at tree level $\mat{M}$ is diagonal.
Off-diagonal elements in $\mat{M}$ are therefore of second order in
$\mat{H}$, $\lambda$.  At one loop we can then write,
\begin{equation}\label{mphi}
  \delta m_\phi^2 = \delta Z_{m_\phi}^{(1)} m_\phi^2 + \sum_{ab}
  \delta Z_{ab}^{(2)} m_{a} m_{b}
+ \delta Z_{m_\phi}^{(3)} \xi^2 
\end{equation}
in both schemes, masses being physical in OS and renormalized ones in
\ms.  Similarly, for the cubic coupling we have,
\begin{equation}\label{xi}
  \xi_\naught = \mu^{\epsilon/2} Z_\xi^{(1)} \xi + \mu^{\epsilon/2}
  \sum_{ab} \delta Z_{\xi,ab}^{(2)} M_{ab},
\end{equation}
where we can set $M_{ab}=0$ for $a\neq b$ at one loop.  At that order,
$\xi$ mixes with $\mat{M}$ through a triangle diagram that does not
depend on $\xi$ or $m_\phi^2$.

The renormalized fermion field can be related to the bare one as
$\mat{\psi}_\naught=\mat{A}\mat{\psi}$, with $\mat{A}$ a complex
$N\times N$ matrix.  It is more convenient to introduce the polar
decomposition of $\mat{A}$ explicitly,
\begin{equation}\label{psi}
  \mat{\psi}_\naught=\mat{U}\mat{Z}^{1/2}\mat{\psi}, \quad
  \mat{U}=e^{\displaystyle -i\mat{\delta U}}, \quad
  \mat{Z}^{1/2} = \mat{1}+\frac{1}{2}\mat{\delta Z}.
\end{equation}
Both $\mat{\delta U}$, $\mat{\delta Z}$ are hermitian.  It is not
difficult to show (see appendix \ref{sec:appa}), however, that we can
always parametrize a unitary matrix $\mat{U}$ in a neighborhood of the
identity as $\mat{U}=e^{-i\mat{\delta U^\prime}} e^{-i\mat{\delta
    \widetilde{U}}}$, with $\mat{\delta U^\prime}$ a linear
combination of diagonal generators and $\mat{\delta \widetilde{U}}$ of
the remaining ones.  The effect of $\mat{\delta U^\prime}$ is a
flavor-dependent phase reparametrization of the fermion fields.
Although $\LL$ is not invariant under such transformations, physical
quantities (such as $S$-matrix elements) remain unaffected by them.
We choose to set $\mat{\delta U^\prime}=0$.  In fact, as we shall see
below, in OS renormalization conditions do not determine the diagonal
elements of $\mat{\delta U}$.  (Thus, in (\ref{psi}) we have
$\mat{\delta Z}\in \mat{u}(N)$, whereas $\mat{\delta U} \in
\mat{u}(N)/\mat{a}$, with $\mat{u}(N)$ the Lie algebra of $N\times N$
hermitian matrices, and $\mat{a}$ its Cartan subalgebra of diagonal
generators.)

The renormalization matrices $\mat{U}$ and $\mat{Z}^{1/2}$ have
different roles in the theory.  If, for instance, we take $\psi_a$ to
be scalar fields and set $\lambda=0$ in $\LL$, $\mat{Z}^{1/2}$ turns
out to be finite, as we would expect of wave-function renormalization
in a superrenormalizable theory \cite{col2}.  On the other hand, in
OS, mass renormalization contributions make $\mat{U}$ divergent.  If,
furthermore, the scalar fields are real, $\mat{Z}$ must be real
symmetric and $\mat{U}$ orthogonal.  In this case $i \mat{\delta U}$
is antisymmetric, with diagonal elements vanishing by definition,
which is consistent with the fact that no phase redefinitions are
possible for real fields.

The fermion mass matrix mixes under renormalization with $\xi$.
Following the parametrization given in appendix \ref{sec:appb}, we
write,
\begin{equation}\label{mctt}
  \mat{M}_\naught = \mat{U}^\dagger_m \left( \mat{M} +
    \mat{\overline{\delta} M} \right) \mat{U}_m + \mat{\delta
    Z}_m^{(\xi)} \xi,
  \quad
  \left[\mat{\overline{\delta} M},\mat{M} \right] = \mat{0},
  \quad
  \mat{U}_m = e^{\displaystyle -i \mat{\delta U}_m},
\end{equation}
with $\mat{\overline{\delta} M}, \mat{\delta U}_m, \mat{\delta
  Z}_m^{(\xi)}$ hermitian.  Here we have conventionally chosen an
additive parametrization for the first term in $\mat{M}_\naught$.
Since we treat $\xi$ as a perturbative parameter, we collect all
dependence on it in the second term.  At one loop $\mat{\delta
  Z}_m^{(\xi)}=0$, so we can write $\mat{M}_\naught =
\mat{M}+\mat{\delta M}$ with $\mat{\delta M} = \mat{\overline\delta M}
+ i \left[ \mat{\delta U}_m, \mat{M}\right]$.  $\mat{\delta U}_m$ is
only determined up to addition of a matrix commuting with $\mat{M}$.

Finally, the matrix of Yukawa couplings $\mat{H}$ is multiplicatively
renormalized.  The most general multiplicative transformation that can
be applied to $\mat{H}$ preserving hermitianity is of the form (see 
appendix \ref{sec:appb}),
\begin{equation}\label{hnot}
  \mat{H}_\naught=\mu^{\epsilon/2} \mat{W}^\dagger \mat{Z}_H \mat{H}
  \mat{W}, 
  \quad
  \left[ \mat{Z}_H,\mat{H} \right]=0,
  \quad
  \mat{W} = e^{\displaystyle -i\mat{\delta W}},
\end{equation}
with $\mat{Z}_H$, $\mat{\delta W}$ hermitian.  In \ms $\mat{Z}_H$ and
$\mat{W}$ can depend only on $\mat{H}$, $\lambda$ and $d$, so that
$\mat{W}$ commutes with both $\mat{Z}_H$ and $\mat{H}$, and drops from
eq.\ (\ref{hnot}).  This would not be the case, though, if there were
another dimensionless matrix in $\LL$ on which $\mat{W}$ could depend
such as, e.g., another mixing matrix for an additional family of
fermions.

Substituting these definitions into (\ref{lag}) we see that, out of
the three unitary matrices we have introduced, only two combinations
enter $\LL$, namely, $\mat{U}_m \mat{U}$ and $\mat{W} \mat{U}$.  We can
therefore always choose $\mat{U}=\mat{1}$.  In OS, however, we
parametrize the theory so that $\mat{M}_\naught$ and \mat{M} are
simultaneously diagonal by setting $\mat{U}_m=\mat{1}$.

The Lagrangian can then be written in terms of renormalized fields and
counterterms as,
\begin{equation}\label{lag1}
  \begin{split}
    {\cal L} &= 
    -\frac{1}{2} \phi(\Box+m_\phi^2)\phi -\frac{1}{2}\delta Z_\phi
    \phi\Box\phi - \frac{1}{2} \Delta m_\phi^2 \phi^2\\
    &\quad 
    + \overline{\mat{\psi}} (i \dirac - \mat{M}) \mat{\psi}
    + \overline{\mat{\psi}} \mat{\delta Z} i\dirac \mat{\psi}
    - \overline{\mat{\psi}} \mat{\Delta M} \mat{\psi}\\
    &\quad 
    + \mu^{\epsilon/2} \overline{\mat{\psi}} \left( \mat{H} +
    \mat{\Delta H} \right) \mat{\psi} \phi\\
  &\quad
  -\frac{1}{3!} \mu^{\epsilon/2} (\xi+\Delta\xi) \phi^3
  -\frac{1}{4!} \mu^{\epsilon} (\lambda+\Delta\lambda) \phi^4,
  \end{split}
\end{equation}
with the one-loop counterterms defined as,
\begin{subequations}\label{lag2}
  \begin{align}
  \Delta m_\phi^2 &= \delta Z_\phi m_\phi^2 + \delta m_\phi^2~;
  \qquad  
  \Delta\lambda  = \delta \lambda + 2 \lambda\delta Z_\phi\\
  \Delta\xi &= \delta\xi+\frac{3}{2} \xi \delta Z_\phi~;
  \qquad
  \delta\xi = \delta Z_\xi^{(1)} \xi + \sum_{a} \delta
  Z_{\xi,a}^{(2)} m_{a}\\
  \mat{\Delta M} &= \frac{1}{2} \left\{\mat{M},\mat{\delta Z}\right\} +
  \mat{\overline{\delta} M} + i \left[\mat{\delta U} + \mat{\delta
      U}_m, \mat{M}\right] \label{osdm}\\
  \mat{\Delta H} &= \mat{\delta Z}_H \mat{H} + i \left[ \mat{\delta U}
  + \mat{\delta W}, \mat{H}\right] + \frac{1}{2} \left\{ \mat{\delta
  Z}, \mat{H}\right\} + \frac{1}{2} \delta Z_\phi \mat{H} \label{ctyk}.
  \end{align}
\end{subequations}
In order to compute the renormalization parts of the theory we have
first to fix a renormalization scheme.

\subsection{Overspecified counterterm parametrization}
\label{sec:overspec}

In (\ref{mctt}) and (\ref{hnot}) we have given the relation between
bare and renormalized hermitian matrices using the parametrization in
appendix \ref{sec:appb}, which is applicable to generic normal
matrices.  In the particular case of hermitian matrices a more obvious
multiplicative parametrization is a congruence transformation
$\mat{H}_\naught = \mat{A}^\dagger \mat{H} \mat{A}$, with \mat{A}
non-singular and dependent on \mat{H}.  Such parametrization must
satisfy constraint relations, since it is overspecified in the
following sense. 

Let $\mat{f}: \mat{u}(N) \longrightarrow \mat{u}(N)$ be a map of
hermitian matrices, given by $\mat{f}(\mat{H}) = \mat{A}^\dagger
\mat{H} \mat{A}$, with $\mat{A}=\mat{A}(\mat{H})$ non-singular.  Then,
we can find a non-singular $\mat{B}=\mat{B}(\mat{H})$ such that,
$\mat{f}(\mat{H}) = \mat{B}^\dagger \mat{H} \mat{B}$ and,
  \begin{equation}\label{const}
  \left[ \mat{B}\mat{B}^\dagger,\mat{H}\right] = 0,
  \qquad
  \left[ \mat{B}^\dagger\mat{B}, \mat{f}(\mat{H})\right] = 0.
  \end{equation}
To see this, we notice that according to appendix \ref{sec:appb} there
must be a hermitian matrix $\mat{Z}_H$ commuting with \mat{H} and a
unitary matrix \mat{U} such that $\mat{f}(\mat{H})=\mat{U}^\dagger
\mat{Z}_H \mat{H} \mat{U}$.  Defining $\mat{B} = \mat{Z}_H^{1/2}
\mat{U}$, we obtain the result.  The parametrization given in Appendix
\ref{sec:appb} is minimal in the sense that eqs.\ (\ref{const}) are
satisfied identically.

\subsection{On-shell scheme}
\label{sec:OS}
In OS we set $\mat{U}_m=\mat{1}$.  If $\mat{M}$ is regular and
diagonal, and $[\mat{M},\mat{\overline\delta M}]=0$, then
$\overline\delta \mat{M}$ must be diagonal, as well as
$\mat{M}_\naught$.

The renormalization conditions we impose on the scalar field
self-energy $\Pi_\phi (p^2)$ are $\Pi_\phi (m_\phi^2) =0=
\Pi^\prime_\phi (m_\phi^2)$, where the prime stands for
$\partial/\partial p^2$.  Defining,
\begin{equation}\label{ome}
  \Omega_\phi (p^2) = -\frac{1}{8\pi^2} \sum_{a,b} H_{ab} H_{ba}
  \left( m_a^2 a_0(m_a^2) + m_b^2 a_0(m_b^2) + \left((m_a+m_b)^2 - p^2
    \right) b_0(p^2,m_a^2,m_b^2)\right),
\end{equation}
which is the ${\cal O} (\epsilon^0)$ part of the unrenormalized $\phi$
self-energy, we obtain,
\begin{subequations}
\begin{align}
  \Pi_\phi (p^2) &= \Omega_\phi (p^2)-\Omega_\phi (m_\phi^2) -
  (p^2-m_\phi^2) \Omega_\phi^\prime (m_\phi^2) \label{aaa}\\
  \delta Z_\phi &= -\frac{1}{4\pi^2\epsilon} \mathrm{Tr}(\mat{H}^2) +
  \Omega_\phi^\prime (m_\phi^2) \label{aab}\\
  \delta m_\phi^2 &= \frac{\xi^2}{16\pi^2\epsilon} -
  \frac{1}{2\pi^2\epsilon} \left(2 \mathrm{Tr}(\mat{H}^2\mat{M}^2) +
  \mathrm{Tr}\left((\mat{HM})^2\right)\right) -
  \Omega_\phi(m_\phi^2) + 
  m_\phi^2 \Omega_\phi^\prime(m_\phi^2). \label{aac}
\end{align}
\end{subequations}
In (\ref{ome}), $a_0$ and $b_0$ refer to finite parts of loop
integrals, defined in appendix \ref{sec:appc}.  (All dependence on the
renormalization scale $\mu$ is through these integrals and, as is easy
to check, $\Pi_\phi (p^2)$ does not depend on $\mu$ as it should in
OS.)  All masses in this section are physical (pole) masses.

For the fermion two-point function, which is a matrix in flavor space,
we write,
\begin{subequations}\label{aad}
\begin{align}
  \mat{\Gamma} &= \pirac\mat{1} - \mat{M} - \mat{\Pi}(p), &
  \mat{\Pi}(p) &= \pirac \mat{\Sigma}^V(p^2) + \mat{\Sigma}^S(p^2),\\
  \intertext{where the form factors are given by,}
  \mat{\Sigma^V}\negthickspace (p^2) &=
  -\frac{1}{16\pi^2\epsilon} \mat{H}^2 - \mat{\delta Z} +
  \mat{\Omega}^V(p^2), & \mat{\Sigma^S}\negthickspace (p^2) &=
  -\frac{1}{8\pi^2\epsilon} \mat{HMH} + \mat{\Delta M} +
  \mat{\Omega}^S(p^2), \intertext{with,} \Omega_{ab}^V(p^2) &=
  \frac{1}{16\pi^2} \sum_c H_{ac} H_{cb} b_-(p^2,m_\phi^2,m_c^2), &
  \Omega_{ab}^S(p^2) &= \frac{1}{16\pi^2} \sum_c H_{ac} H_{cb} m_c
  b_0(p^2,m_\phi^2,m_c^2).
\end{align}
\end{subequations}
Renormalization conditions are expressed in terms of
$\mat{\Sigma}^{V,S}(p^2)$ as \cite{den},
\begin{subequations}
\begin{align}
    m_a \Sigma_{aa}^V(m_a^2) + \Sigma_{aa}^S(m_a^2) &= 0, &
    \Sigma_{aa}^V(m_a^2) + 2 m_a^2 \Sigma_{aa}^{V\prime}(m_a^2) + 2
    m_a \Sigma_{aa}^{S\prime}(m_a^2) &= 0 \label{osrc1}\\
    m_b \Sigma_{ab}^V(m_b^2) + \Sigma_{ab}^S(m_b^2) &= 0, &
    m_a \Sigma_{ab}^V(m_a^2) + \Sigma_{ab}^S(m_a^2) &= 0, \label{osrc2}
\end{align}
\end{subequations}
where in the second line $a\neq b$.  With these conditions, 
we obtain for the diagonal two-point functions,
\begin{subequations}\label{OSig}
\begin{align}
  \Sigma_{aa}^V(p^2) &= \Omega_{aa}^V(p^2)-\Omega_{aa}^V(m_a^2) - 2
  m_a^2 \Omega_{aa}^{V\prime}(m_a^2) - 2 m_a
  \Omega_{aa}^{S\prime}(m_a^2) \\
  \Sigma_{aa}^S(p^2) &= \Omega_{aa}^S(p^2)-\Omega_{aa}^S(m_a^2) + 2
  m_a^3 \Omega_{aa}^{V\prime}(m_a^2) + 2 m_a^2
  \Omega_{aa}^{S\prime}(m_a^2),
\end{align}
and for the off-diagonal ones, ($a\neq b$)
\begin{align}
  \Sigma_{ab}^V(p^2) &= \Omega_{ab}^V(p^2) - \frac{m_a
  \Omega_{ab}^V(m_a^2) - m_b \Omega_{ab}^V(m_b^2)}{m_a - m_b} -
  \frac{\Omega_{ab}^S(m_a^2) - \Omega^S_{ab}(m_b^2)}{m_a-m_b}\\
  \Sigma_{ab}^S(p^2) &= \Omega_{ab}^S(p^2) + \frac{m_a m_b}
  {m_a - m_b} \left(\Omega_{ab}^V(m_a^2) - \Omega_{ab}^V(m_b^2)\right)
  + \frac{m_b \Omega_{ab}^S(m_a^2) - m_a
  \Omega^S_{ab}(m_b^2)}{m_a-m_b}. 
\end{align}
\end{subequations}
In both cases, diagonal and off-diagonal, renormalization constants
can be compactly expressed as,
\begin{equation}\label{osctt}
  \mat{\delta Z} = -\frac{1}{16\pi^2\epsilon} \mat{H}^2 +
  \mat{\Omega}^V(p^2) - \mat{\Sigma}^V(p^2)~;
  \qquad
  \mat{\Delta M} = \frac{1}{8\pi^2\epsilon} \mat{HMH} -
  \mat{\Omega}^S(p^2) + \mat{\Sigma}^S(p^2)~. 
\end{equation}
It is immediate to check, by substituting the expressions for
$\mat{\Sigma}^{V,S}$ into these equations, that $\mat{\delta Z}$ and
$\mat{\Delta M}$ are hermitian.  With this, the fermion propagator is
renormalized to one loop.  We can, however, compute the mass
counterterm $\mat{\overline\delta M}$ and $\mat{\delta U}$ from the
expressions (\ref{osctt}) for $\mat{\Delta M}$ and $\mat{\delta Z}$.
From (\ref{osdm}) we have,
\begin{equation}
  \mat{\overline\delta M} + i [\mat{\delta U}, \mat{M}] = \mat{\Delta
  M} - \frac{1}{2} \{\mat{\delta Z}, \mat{M}\}.\label{fitt}
\end{equation}
This equation can always be solved for $\mat{\overline\delta M}$ and
$\mat{\delta U}$, as discussed in appendix \ref{sec:kernel}.  In this
case, in which $\mat{M}$ is diagonal and regular, however,
(\ref{fitt}) is trivial to solve.  $\mat{\overline\delta M}$ is given
by the diagonal elements of the rhs of (\ref{fitt}) and $\mat{\delta
  U}$ by the off-diagonal ones,
\begin{subequations}
  \begin{align}
    \overline\delta M_{ab} &= (\Delta M_{aa}-m_a \delta Z_{aa})
    \delta_{ab} \label{osdmbar}\\
    \delta U_{ab} &= \frac{i}{m_a-m_b} \left(\Delta M_{ab} -
      \frac{m_a+m_b}{2} \delta Z_{ab}\right);
    \qquad a\neq b.\label{osu}
  \end{align}
\end{subequations}
The diagonal elements $\delta U_{aa}$ are not determined by
renormalization conditions.  As mentioned above, they only change the
phases of fermion fields, and can be chosen to vanish.

We consider next the renormalization of the Yukawa coupling, given by
the 1-PI three-point Green's function $\mat{\Gamma}(p_1,p_2)$.  (Here
$p_1$ is the momentum of the incoming scalar and $p_{2,3}$ are the
momenta of the outgoing fermions.)  At tree level
$\mat{\Gamma}=\mat{H}$ is a Lorentz scalar.  Expanding
the one-loop $\mat{\Gamma}$ in the usual $\gamma$-matrix basis, it is
clear that only the scalar form-factor can receive divergent
contributions, since the counterterm for $\mat{\Gamma}$ in $\LL$ is a
scalar, the other form factors being finite.  This is easily seen also
from the explicit form of the corresponding loop integrals, since by
power counting only those terms containing two powers of loop momentum
in the numerator are divergent, and they contribute to the scalar form
factor only.

We will then focus on $\mat{F}(p_1,p_2)\equiv 1/4
\tr(\mat{\Gamma}(p_1,p_2))$.  For concreteness, we assume that
the physical value of the coupling $\mat{H}$ is fixed by this form
factor.  \mat{F} is a scalar function of momenta, depending on $p_1$,
$p_2$ only through $p_1^2$, $p_2^2$, $(p_1-p_2)^2$.  These
combinations are fixed if the external momenta are required to be on
their mass shell.  Calling $\os{F_{ab}}$ the value of
$F_{ab}(p_1,p_2)$ at $p_1^2 = m_\phi^2$, $p_2^2 = m_a^2$,
$(p_1-p_2)^2 = m_b^2$, we impose the renormalization condition
\begin{equation}\label{yukren}
  \os{\mat{F}} = \mat{H}.
\end{equation}
(We notice, furthermore, that the hermitianity of the lhs is
guaranteed by CPT invariance, which also requires \mat{H} to be
hermitian.)  At one loop $\mat{F}(p_1,p_2)$ is given by
\begin{align}
    F_{ab}(p_1,p_2) &= \mu^{\epsilon/2} (H_{ab}+\Delta H_{ab})
    -\mu^{\epsilon/2} \frac{1}{8\pi^2\epsilon}
    \left(\mat{H}^3\right)_{ab}
    +\mu^{\epsilon/2} \sum_{c,d} H_{ac} H_{cd} H_{db}
    h_1(p_1,p_2;m_\phi^2,m_c^2,m_d^2)\notag\\
    &\quad -\mu^{\epsilon/2} \xi \sum_c H_{ac} H_{cb}
    h_2(p_1,p_2;m_\phi^2,m_c^2),
\end{align}
where the loop integrals $h_{1,2}$ are defined in appendix
\ref{sec:appc}. 
The renormalization condition (\ref{yukren}) then leads to,
\begin{equation}\label{osyct}
  \Delta H_{ab} = \frac{1}{8\pi^2\epsilon} \left(\mat{H}^3\right)_{ab}
  - \sum_{cd} H_{ac} H_{cd} H_{db} \os h_1(m_\phi^2,m_c^2,m_d^2) +
  \xi \sum_c H_{ac} H_{cb} \os h_2(m_\phi^2,m_c^2),
\end{equation}
where $\os h_{1,2}$ refer to the on-shell values of those integrals.
This counterterm is enough to renormalize the scalar form-factor at
one loop.  We can, in principle, determine the counterterms to the
coupling matrix $\mat{H}$ as defined in (\ref{ctyk}) through the
equation,
\begin{equation}\label{fith}
  \mat{\delta Z}_H \mat{H} + i \left[ \mat{\delta W}, \mat{H}\right] =
  \mat{\Delta H} - i \left[ \mat{\delta U} , \mat{H}\right] -
  \frac{1}{2} \left\{ \mat{\delta Z}, \mat{H}\right\} - \frac{1}{2}
  \delta Z_\phi \mat{H},
\end{equation}
where all the quantities on the rhs are already known.  Equation
(\ref{fith}) always has a solution, since by definition $[\mat{\delta
  Z}_H,\mat{H}]=0$, as shown in appendix \ref{sec:kernel}.  Unlike
(\ref{fitt}), in this case it is more difficult to find the solution
algebraically in closed form, only some of the contributions to each
counterterm being obvious,
\begin{equation}
  \mat{\delta Z}_H = \frac{1}{8\pi^2\epsilon}
  \left(\mat{H}^2\right) - \frac{1}{2}
  \delta Z_\phi + \cdots~;
  \qquad
  \left[ \mat{\delta W}, \mat{H}\right] = -  \left[ \mat{\delta U} ,
  \mat{H}\right] + \cdots .
\end{equation}
Eq.\ (\ref{fith}) can be solved by projecting it over an appropriate
basis for the algebra $\mat{u}(N)$ (see appendix \ref{sec:kernel}),
which can be done numerically.  It should be clear, however, that
counterterms are completely fixed by renormalization conditions.  Once
those are established, no other choices are involved.  We notice that
diagonal elements of $\mat{\delta U}$ contribute to $\mat{\delta W}$,
as expected, since a change of phase of fermion fields leads to a
corresponding change in $\mat{H}$.

\subsection{The mass-degenerate case in on-shell scheme}
\label{sec:degenerate}

As discussed in \S \ref{sec:OS} in OS we set $\mat{U}_m=\mat{1}$, so
that $ \mat{M}_\naught = \mat{M} + \mat{\overline\delta M},$ with
$[\mat{M},\mat{\overline\delta M}] = 0.$ $\mat{M}_\naught$ and \mat{M}
can then be simultaneously diagonalized.  If \mat{M} is regular,
choosing a basis in which it is diagonal immediately implies that
$\mat{\overline\delta M}$ is diagonal, and then so is
$\mat{M}_\naught$.  If \mat{M} is degenerate, though,
$\mat{\overline\delta M}$ need not be diagonal even if \mat{M} is.  We
notice also that \mat{H} can be diagonalized within each eigenspace of
\mat{M} by means of a unitary transformation of ${\cal O}(\mat{H}^0)$.
We assume that such a transformation has been performed and then, in
general, $\mat{\overline\delta M}$ will not be diagonal within mass
eigenspaces.

For notational simplicity we assume that there is one degenerate mass
eigenvalue, say $m_1$, with multiplicity $1<r<N$ ($r=N$ being the
trivial case $\mat{M}\propto \mat{1}$), all other masses, $m_{r+1},
\ldots, m_N,$ being non-degenerate.  The generalization to the case of
several degenerate eigenvalues presents no difficulties.

The renormalized fermion two-point Green's functions, as given by
(\ref{OSig}) and (\ref{osctt}) are unchanged in the mass-degenerate
case if $a=b$ or if $a\neq b$ but $\max(a,b)>r$.  If $a\neq b$ and
both $a,b\leq r$, those results need modification.

Renormalization conditions (\ref{osrc2}) reduce to the single equation
($a\neq b$, $a,b\leq r$)
\begin{equation}
  \label{osrc3}
   m_1 \Sigma_{ab}^V(m_1^2) + \Sigma_{ab}^S(m_1^2) = 0,
\end{equation}
leading to,
\begin{equation}\label{osrc4}
  \Delta M_{ab} - m_1 \delta Z_{ab} = \frac{m_1}{16\pi^2\epsilon}
  \left(\mat{H}^2\right)_{ab} + \frac{1}{8\pi^2\epsilon}
  (\mat{HMH})_{ab} - m_1 \Omega^V_{ab}(m_1^2) - \Omega^{S}_{ab}(m_1^2).
\end{equation}
From definition (\ref{osdm}) we see that in this case, $\Delta M_{ab}
- m_1 \delta Z_{ab} = \overline\delta M_{ab}$.  Thus, the
renormalization condition (\ref{osrc3}) fixes the off-diagonal
counterterm $\overline\delta M_{ab}$ to the value given in
(\ref{osrc4}), and that is all that is required to renormalize the
two-point function within the eigenspace of $m_1$.  $\overline\delta
M_{ab}$, as given by (\ref{osdmbar}) when $a=b$ or $a\neq b$,
$\max(a,b)>r$, and by (\ref{osrc4}), when $a\neq b$ and $a,b \leq r$,
obviously commutes with $\mat{M}$, since it is non-diagonal only
within eigenspaces of $\mat{M}$.  Furthermore, we set $\delta
U_{ab}=0$, for $a,b \leq r$, instead of (\ref{osu}).  Notice that when
$r=N$, \mat{H} is diagonal and therefore so are $\mat{\Omega}^{V,S}$
and \mat{\overline\delta M}.

If, however, we want to make each form factor $\Sigma_{ab}^{V,S}$
finite separately, we may impose additional renormalization
conditions.  These are quite arbitrary, as long as they are consistent
with (\ref{osrc3}) (or (\ref{osrc4})).  We can, for instance, take the
limit of degenerate masses in (\ref{OSig}) and (\ref{osctt}) to get,
for $a\neq b$, $a,b \leq r$, 
\begin{subequations}\label{extr1}
\begin{align}
  \delta Z_{ab} &= -\frac{1}{16\pi^2\epsilon} (\mat{H}^2)_{ab} +
  \Omega^V_{ab}(m_1^2) + 2 m_1^2 \Omega^{V\prime}_{ab}(m_1^2) + 2 m_1
  \Omega^{S\prime}_{ab}(m_1^2) \\
  \Delta M_{ab} &= \frac{1}{8\pi^2\epsilon} (\mat{HMH})_{ab} -
  \Omega^S_{ab}(m_1^2) + 2 m_1^3 \Omega^{V\prime}_{ab}(m_1^2) + 2 m_1^2
  \Omega^{S\prime}_{ab}(m_1^2). 
\end{align}
\end{subequations}
Requiring instead $\Sigma^{V,S}_{ab}(m_1^2)=0$, leads to,
\begin{equation}\label{extr2}
  \delta Z_{ab} =  -\frac{1}{16\pi^2\epsilon} (\mat{H}^2)_{ab} +
  \Omega^V_{ab}(m_1^2) ~;
  \qquad
  \Delta M_{ab} = \frac{1}{8\pi^2\epsilon} (\mat{HMH})_{ab} -
  \Omega^S_{ab}(m_1^2).   
\end{equation}
Both (\ref{extr1}) and (\ref{extr2}) are consistent with
(\ref{osrc3}).

\subsection{\ms scheme}
\label{sec:msbar}

In \ms we set $\mat{U}=\mat{1}$. We choose a flavor basis for fermion
fields so that at tree level the renormalized mass matrix \mat{M} is
diagonal, $\mat{M}=\mat{M}_\mathrm{phys}$.  Off-diagonal elements of
\mat{M} are then ${\cal O}(\mat{H}^2)$.  We write
$\mat{M}=\mat{M}^\prime + \widehat{\mat{M}}$, with
$\mat{M^\prime}=\mathrm{diag}(m_1,\ldots,m_N)$ containing the
renormalized masses and $\widehat{\mat{M}}$ the off-diagonal elements,
and treat $\widehat{\mat{M}}$ as an interaction term.  We write the
fermion Lagrangian as,
\begin{align*}
  \cal{L}_\psi &= \overline{\mat{\psi}} (i\dirac - \mat{M}^\prime)
  \mat{\psi} - \overline{\mat{\psi}} \widehat{\mat{M}}\mat{\psi} +
  \overline{\mat{\psi}} (\mat{\delta Z} i\dirac - \mat{\Delta M})
  \mat{\psi} \\
  \mat{\Delta M} &= \mat{\overline\delta M} + i [\mat{\delta U}_m,
  \mat{M}] + \frac{1}{2} \{\mat{M}, \mat{\delta Z}\},
\end{align*}
instead of the second line of (\ref{lag1}).  The tree-level fermion
propagator is thus flavor diagonal.

We notice, parenthetically, that \mat{M} can be written as $\mat{M} =
\exp(i\mat{E}) \mat{M}' \exp(-i\mat{E})$, with $\mat{M}'$ the diagonal
matrix of eigenvalues and $\exp(-i\mat{E})$ the unitary matrix of
eigenvectors of \mat{M}.  Our choice of tree-level flavor basis
implies $\mat{E}={\cal O}(\mat{H}^2)$.  At one-loop level, then,
$\mat{M}'$ is given by the diagonal entries of \mat{M} and
$\widehat{\mat{M}} = i [\mat{E},\mat{M}']$ contains the off-diagonal
ones.

Counterterms for the scalar two-point function, as defined by
(\ref{lag1}) and (\ref{lag2}), are given by the $\epsilon$-pole terms
of (\ref{aab}) and (\ref{aac}).  The $\phi$ self-energy is then
$\Pi_\phi (p^2) = \Omega_\phi (p^2)$, with $\Omega_\phi$ defined in
(\ref{ome}).  Requiring that the two-point function have a zero at
$p^2=m_{\phi \mathrm{ph}}^2$ leads to the relation $m_{\phi}^2 =
m_{\phi \mathrm{ph}}^2 - \Omega_\phi(m_{\phi \mathrm{ph}}^2)$ at one
loop.

The fermion two-point function can also be read off the corresponding
OS results.  Starting from (\ref{aad}), we set \mat{\delta Z} and
\mat{\Delta M} to be the $\epsilon$ poles of (\ref{osctt}), to obtain
$\mat{\Sigma}^{V,S} (p^2) = \mat{\Omega}^{V,S} (p^2)$.  The
counterterms \mat{\overline\delta M} and $\mat{\delta U}_m$ can then
be obtained from \mat{\delta Z} and \mat{\delta M} in the same way as
in OS (see (\ref{fitt})).  In this case, unlike in OS, they are not
needed to renormalize the Yukawa coupling.  We define
\begin{equation}
  \label{deltam}
  \mat{\delta M} = \mat{\Delta M} - \frac{1}{2} \{\mat{M},\mat{\delta
    Z}\} = \frac{1}{8\pi^2 \epsilon} \left( \mat{HMH} + \frac{1}{4}
    \left\{ \mat{M},\mat{H}^2 \right\}\right),
\end{equation}
which we will use in the one-loop renormalization group (RG) equation
for \mat{M} below.

We require $\Gamma_{aa}(p) u_a(p)|_{p^2=m_{a\mathrm{ph}}^2} =0$ in
order to obtain renormalized masses in terms of physical ones, leading
to the relation
\begin{equation}
  \label{massa}
  m_a = m_{a\mathrm{ph}} - \left( m_{a\mathrm{ph}}
  \Omega^{V}_{aa}(m_{a\mathrm{ph}}^2) +
  \Omega^{S}_{aa}(m_{a\mathrm{ph}}^2) \right).
\end{equation}
Expressing the off-diagonal elements $\widehat{M}_{ab}$, $a\neq b$, in
terms of physical masses and coupling constants involves a choice of
parametrization of the theory.  In principle, any value of
$\widehat{\mat{M}}$ consistent with the renormalization group
equations (given below) is admissible.  Different choices of
$\widehat{\mat{M}}$ at one-loop level will result in different
parametrizations of the renormalized masses $m_a$ (the eigenvalues of
\mat{M}), in terms of the physical ones $m_{a\mathrm{ph}}$ at two
loops.  We could give $\widehat{\mat{M}}$ a definite value at a mass
scale $\overline\mu_\naught$, and evolve it with the RG equations to
the desired scale $\overline\mu$.  For concreteness, we quote the
expression for \mat{M} obtained by matching the theory in
\ms to the OS results,
\begin{equation}\label{match}
  M_{ab} = (\mat{M}_{\mathrm{ph}})_{ab} - \frac{1}{2} \left(
    m_{a\mathrm{ph}} \Omega^V_{ab} (m_{a\mathrm{ph}}^2) +
    \Omega^S_{ab} (m_{a\mathrm{ph}}^2) +
    m_{b\mathrm{ph}} \Omega^V_{ab} (m_{b\mathrm{ph}}^2) +
    \Omega^S_{ab} (m_{b\mathrm{ph}}^2) \right),
\end{equation}
where $\mat{M}_\mathrm{ph} = \mathrm{diag}(m_{1\mathrm{ph}},\ldots,
m_{N\mathrm{ph}})$.  When $a=b$ this equation reduces to
(\ref{massa}).  

It is convenient to discuss at this point the effect of a finite,
unitary renormalization of fermion fields, with $\mat{\delta U}= {\cal
  O}(\mat{H}^2)$ and $d$-independent. (We exclude from consideration
unitary transformations with $\mat{\delta U} = {\cal O}(\mat{H}^0)$,
which correspond to transformations of the classical fields.)  The
effect of such a transformation on the basic Lagrangian is to change
$\mat{M} \rightarrow \mat{U}^\dagger \mat{M} \mat{U}$ and $\mat{H}
\rightarrow \mat{U}^\dagger \mat{H} \mat{U}$.  It is clear that
$\mat{M}$ remains diagonal at ${\cal O}(\mat{H}^0)$, so our choice of
a tree-level flavor basis is not altered.  Also, $\mat{M}^\prime$
remains unchanged through ${\cal O}(\mat{H}^2)$, so (\ref{massa})
still holds.  Changes in $\mat{M}^\prime$ start at ${\cal
  O}(\mat{H}^4)$.  On the other hand, $\widehat{M}$ changes by a term
$i[\mat{\delta U},\mat{M}^\prime]$ at ${\cal O}(\mat{H}^2)$.
Counterterms still are of the \ms form after the transformation, which
is multiplicative and independent of $\epsilon$.  (If, however,
\mat{\delta U} depends on masses, counterterms acquire mass
dependence.)  We see, then, that this $U(N)$ freedom to perform
finite unitary renormalizations of \mat{\psi} is a source of
arbitrariness in $\widehat{\mat{M}}$.  By choosing a definite value
for $\mat{M}$, as in (\ref{match}), we are reducing the ambiguity in
the choice of one-loop flavor basis for \mat{\psi} to the subgroup of
$U(N)$ that leaves \mat{M} invariant.  If \mat{M} is regular, this is
the $(U(1))^N$ Abelian subgroup of flavor-dependent phase
transformations.

The relation between bare and renormalized Yukawa couplings is given
by (\ref{hnot}).  As discussed in relation to that equation, the
unitary renormalization matrix \mat{W} is trivial in \ms in this
model.  From the value of \mat{\Delta H} in OS, eq.\ (\ref{osyct}),
we get
\begin{equation*}
  \mat{\Delta H} = \frac{1}{8\pi^2\epsilon} \mat{H}^3.
\end{equation*}
Therefore, from definition (\ref{ctyk}) we obtain,
\begin{equation*}
  \mat{\delta Z}_H = \frac{1}{16\pi^2\epsilon}
  (3\mat{H}^2+2\tr(\mat{H}^2)).
\end{equation*}
With this value for $\mat{\delta Z}_H$ we can immediately find the
one-loop \mat{\beta} function for \mat{H},
\begin{equation}
  \label{beta}
  \overline\mu \frac{d\mat{H}}{d\overline\mu} = \mat{\beta}
  = -\frac{\epsilon}{2} \mat{H} + \frac{1}{16\pi^2}
  \left( 3 \mat{H}^3 + 2 \tr(\mat{H}^2) \mat{H} \right) +
  {\cal O}(\mat{H}^5).
\end{equation}
\mat{\beta}, which does not depend on $\lambda$ at one loop, commutes
with \mat{H}.  This need not be the case in more complicated theories
where \mat{\beta} can depend on other dimensionless matrices.

With \mat{\delta M} and \mat{\beta}, from (\ref{deltam}) and
(\ref{beta}), we obtain the evolution equation for \mat{M},
\begin{equation}
  \label{rgmass}
  \overline\mu \frac{d\mat{M}}{d\overline\mu} = -\mat{\gamma}_m
  = \frac{1}{16\pi^2} \left( 2 \mat{HMH} + \frac{1}{2}
  \{ \mat{M}, \mat{H}^2 \}\right) + {\cal O}(\mat{H}^4).
\end{equation}
The matrix $\mat{\gamma}_m$ defined by this equation is unconventional
in that it has mass dimension one.  At one-loop level we can take
\mat{M} on the rhs to be diagonal (in fact, we can set
$\mat{M}=\mat{M}_\mathrm{ph}$) and set \mat{H} to its tree-level
value, neglecting its $\overline\mu$ dependence.  The solution to
(\ref{rgmass}) up to terms of ${\cal O}(\mat{H}^4)$ is then,
\begin{equation}
  \mat{M}(\overline\mu) = \mat{M}(\overline\mu_\naught) -
  \mat{\gamma}_m \log\left( \frac{\overline\mu}{\overline\mu_\naught}
  \right).  
\end{equation}
We see that we can diagonalize the one-loop mass matrix at a given
scale $\overline\mu_\naught$.  If $[\mat{M},\mat{H}]=0$, then \mat{M}
also commutes with $\mat{\gamma}_m$ and we can diagonalize \mat{M} at
all scales.  This is the case, of course, if $\mat{M}\propto \mat{1}$
or $\mat{H}\propto \mat{1}$.  It is easy to check that the expression
(\ref{match}) for \mat{M} is a solution to (\ref{rgmass}).

We consider, finally, the anomalous dimensions of fields.  From the
expressions for $\delta Z_\phi$ and \mat{\delta Z} we find,
\begin{equation}
  \overline\mu\frac{d\phi}{d\overline\mu} = -\frac{1}{8\pi^2}
  \tr\left( \mat{H}^2 \right) \phi ~;
  \qquad
  \overline\mu\frac{d\mat{\psi}}{d\overline\mu} = -\frac{1}{32\pi^2}
  \mat{H}^2 \mat{\psi}.
\end{equation}
Once again, the dependency of \mat{H} on $\overline\mu$ can be
neglected in the rhs of these equations.  The evolution of \mat{\psi}
with $\overline\mu$ is given by an hermitian matrix, which is
consistent with \mat{U}=\mat{1}.

\section{Normal mixing matrix}
\label{sec:normal}

The treatment given in \S \ref{sec:hermitean} for the case of a
hermitian mixing matrix can be readily generalized to normal mixing
matrices.  Such a generalization is natural, since it comprises all
mixing matrices that can be diagonalized by a unitary transformation
of fields.  Consider, for instance, a Yukawa interaction of the form,
\begin{equation} 
  \label{lagnormal}
  \LL = -\phi_\naught^\dagger (\Box+{m_\phi}_\naught^2) \phi_\naught
  + \overline{\mat{\psi}}_\naught (i \dirac - \mat{M}_\naught)
  \mat{\psi}_\naught
  + \overline{\mat{\psi}}_\naught \mat{N}_\naught
  \mat{\psi}_\naught \phi_\naught
    + \overline{\mat{\psi}}_\naught \mat{N}_\naught^\dagger
  \mat{\psi}_\naught \phi_\naught^\dagger
  -V\left(\phi,\phi^\dagger\right),
\end{equation} 
where $V$ contains cubic and quartic terms and
$[\mat{N},\mat{N}^\dagger]=0$.   This interaction induces
$\phi$-$\phi^\dagger$ mixing, which must be taken into account when
renormalizing the theory.  Besides that extra complication, the
renormalization can be carried out along the same lines as in \S
\ref{sec:hermitean}, where now,
\begin{equation}
  \label{eq:norm0}
  \mat{N}_\naught = \mat{U}_N^\dagger \mat{Z}_N \mat{N} \mat{U}_N,
\end{equation}
with $\mat{U}_N$ unitary and $\mat{Z}_N$ normal such that
$[\mat{Z}_N,\mat{N}]=0$.  Unlike the previous case, though, we cannot
parametrize $\mat{N}_\naught$ in the form $\mat{B}^\dagger \mat{N}
\mat{B}$, since congruence transformations do not preserve normality.

More generally, we must ask whether the normality of \mat{N} is stable
under renormalization.  (The hermitianity of \mat{H} in \S
\ref{sec:hermitean} was obviously stable by unitarity.)  That this is
so can be seen by noticing that the Yukawa interaction in
(\ref{lagnormal}) is invariant under a $(U(1))^N$ symmetry which, in a
flavor basis in which \mat{N} is diagonal, is given by $\psi_a
\rightarrow e^{-i \alpha_a} \psi_a$, $\phi\rightarrow \phi$.  This
$(U(1))^N$ symmetry will generally be broken by fermion mass terms.
In \ms the Yukawa coupling will remain normal after radiative
corrections are taken into account, since the fermion wave-function
renormalization matrix is diagonal in the interaction basis due to the
$(U(1))^N$ symmetry (or, in this model, because it must be a
polynomial in $\mat{N}$).  In OS finite asymmetric counterterms to the
Yukawa coupling are needed, which render the renormalized Lagrangian
asymmetric.  An example of this well-known phenomenon (see,
e.g., \cite[Ch.4]{cole} and refs.\ therein) is considered in
the next section in connection with unitary mixing.
\section{Unitary mixing matrix}
\label{sec:unitary}

The case of a unitary mixing matrix constitutes an important example
of non-hermitian, normal mixing.  Besides its obvious phenomenological
relevance \cite{sac,gam,bar}, this case is interesting because mixing
matrix unitarity imposes restrictive constraints on the form of
counterterms, on top of those stemming from normality.  

Furthermore, by considering unitary mixing we can extend our approach
to interactions of the general form $\overline{\mat{\psi}}_1 \mat{V}
\mat{\psi}_2$, with two different families of fermions.  Independent
unitary transformations of the fermion fields lead to a biunitary
transformation of the mixing matrix \mat{V}.  In general normality is
not preserved by biunitary transformations, so our previous results do
not apply to this type of interactions unless \mat{V} is unitary.

In this case, the $(U(1))^N$ flavor symmetry of the previous section
is replaced by a larger $SU(N)$ invariance, since we can choose the
phases of fermion fields so that in the interaction basis the mixing
matrix is just the identity.  For concreteness we consider the simple
Lagrangian, written in the interaction basis,
\begin{align}
  \label{eq:laguni}
    \LL & = \LL_\gamma + \LL_\phi + \LL_\psi + \LL_Y + \LL_\mathrm{em} - V
    \\
    \LL_\gamma &= -\frac{1}{4} F_{\mu\nu} F^{\mu\nu} + \frac{1}{2}
    m_\gamma^2 A_\mu A^\mu - \frac{1}{2\xi} \left(\partial_\mu
      A^\mu\right)^2,  &
    \LL_\phi &= -\phi^\dagger (\Box + m_\phi^2) \phi \notag\\
    \LL_\psi &= \sum_{j=1}^2 \overline{\mat{\psi}}_j \left( i\dirac -
      P_+
      \mat{M}_j - P_- \mat{M}_j^\dagger \right) \mat{\psi}_j, &
    \LL_Y &= g \overline{\mat{\psi}}_1 P_+ \mat{\psi}_2 \phi +
    g \overline{\mat{\psi}}_2 P_- \mat{\psi}_1 \phi^\dagger\notag\\
    \LL_\mathrm{em} &= Q_\phi^2 e^2 \phi^\dagger \phi A_\mu
    A^\mu + \left(-j_\phi^\mu + j_1^\mu + j_2^\mu\right) A_\mu, &
    V&= \frac{\lambda}{6} \left(\phi^\dagger \phi\right)^2 \notag\\
    j_\phi^\mu &= Q_\phi e i \left( \partial^\mu \phi^\dagger \phi -
      \phi^\dagger \partial^\mu \phi  \right), &
    j_j^\mu &= Q_j e \overline{\mat{\psi}}_j\gamma^\mu \mat{\psi}_j. \notag
\end{align}
The model consists of two fermion families $\mat{\psi}_{1,2}$, each
containing $N$ flavors.  These interact with a scalar field $\phi$
trough the Yukawa couplings in $\LL_Y$, which are arbitrarily chosen to
be chiral, with $P_{\pm}=(1\pm \gamma_5)/2$.  We choose $g$ to be real
and positive, since its phase can always be absorbed in $\phi$. All of
these fields are charged, with $Q_\phi=Q_1-Q_2$, and minimally coupled
to a massive photon field $A^\mu$.  $U(1)$ gauge invariance is broken
explicitly by the photon mass and by the covariant gauge fixing terms.
One-loop radiative corrections to the Yukawa interaction arise only
from the couplings of $\mat{\psi}_j$ and $\phi$ to the massive photon.

Besides $U(1)$ gauge invariance, $\LL$ has a global $SU(N)$ flavor
symmetry $ \mat{\psi}_j \rightarrow e^{-i\sum_A \alpha_A
  \mat{\lambda}_A} \mat{\psi}_j$, broken by fermion mass differences,
with associated current,
\begin{equation}
j^\mu_A  = \sum_{j=1}^2 \overline{\mat{\psi}}_j \gamma^\mu
\mat{\lambda}_A \mat{\psi}_j,
\qquad
\partial_{\mu}j^\mu_A = i \sum_{j=1}^2 \overline{\mat{\psi}}_j
\left( P_+ [\mat{M}_j,\mat{\lambda}_A] + P_- [\mat{M}_j^\dagger,
\mat{\lambda}_A]\right) \mat{\psi}_j.
\end{equation}
Here, $\mat{\lambda}_A$ ($A=1,\ldots,N^2-1$) are the generators of
$SU(N)$ in the fundamental representation.  $\LL$ has also broken
global axial $U(1)_5$ and $SU(N)_5$ symmetries which will not be
needed in the sequel.

The renormalization of this model proceeds much in the same way as in
\S \ref{sec:hermitean}.  We introduce renormalization constants
$Z_\phi$, $\delta m_\phi^2$ for the $\phi$ field and its mass. For the
Yukawa coupling we write $g_\naught = \mu^{\epsilon/2} Z_g g$ , and
for the scalar self-coupling, $\lambda = \mu^\epsilon (\lambda +
\delta \lambda$). By gauge invariance, only wave-function
renormalization is needed for the gauge sector \cite{col2,zin},
\begin{equation*}
  A^\mu_\naught = Z_\gamma^{1/2} A^\mu,
  \quad
  m_{\gamma\naught}^2 = Z_\gamma m_{\gamma}^2,
  \quad
  \xi_\naught = Z_\gamma \xi,
  \quad
  e_\naught = \mu^{\epsilon/2} Z_\gamma^{-1/2} e.
\end{equation*}

Henceforth, we work in a flavor basis such that at tree level mass
matrices are diagonal.  We can always find unitary matrices
$\mat{V}_j$, $\mat{W}_j$, ($j=1,2$), so that the mass matrices are
written as $\mat{M}_j=\mat{V}_j \mat{M}_j'\mat{W}_j^\dagger$, with
$\mat{M}_j'$ diagonal, with positive entries \cite[\S 21.3]{swe}.  The
corresponding transformation of fermion fields is $\mat{\psi}_j =
(\mat{W}_j P_+ + \mat{V}_j P_-) \mat{\psi}_j'$.  Writing the
Lagrangian in terms of primed quantities, and dropping the primes, at
tree level we get,
\begin{equation}\label{lab1}
  \LL_\psi  = \sum_{j=1}^2 \overline{\mat{\psi}}_j \left( i\dirac -
    \mat{M}\right) \mat{\psi}_j,
  \qquad
  \LL_Y = g \overline{\mat{\psi}}_1
  P_+ \mat{V}\mat{\psi}_2 \phi +
  g \overline{\mat{\psi}}_2 P_- \mat{V}^\dagger \mat{\psi}_1
  \phi^\dagger,
\end{equation}
with $\mat{V}= \mat{V}_1^\dagger \mat{W}_2$ the unitary mixing matrix.
In this basis, the $SU(N)$ global flavor symmetry is given by,
\begin{subequations}
\begin{align}
  \mat{\psi}_1 &\longrightarrow e^{-i\sum_A \alpha_A \mat{\lambda}_A}
  \mat{\psi}_1, & 
  \mat{\psi}_2 \longrightarrow \mat{V}^\dagger e^{-i\sum_A \alpha_A
    \mat{\lambda}_A} \mat{V} \mat{\psi}_2 \\
    j^\mu_A & = \overline{\mat{\psi}}_1 \gamma^\mu
  \mat{\lambda}_A \mat{\psi}_1 + \overline{\mat{\psi}}_2 \gamma^\mu
  \mat{V}^\dagger \mat{\lambda}_A \mat{V} \mat{\psi}_2, &
  \partial_{\mu}j^\mu_A = i \overline{\mat{\psi}}_1
  [\mat{M}_1,\mat{\lambda}_A] \mat{\psi}_1 +
  i \overline{\mat{\psi}}_2
  [\mat{M}_2,\mat{V}^\dagger \mat{\lambda}_A \mat{V}] \mat{\psi}_2. 
\end{align}
\end{subequations}
Notice that we can parametrize this flavor symmetry, and its
associated currents, in an infinite number of different ways.  We
could have written, for example, $\mat{\psi}_j \rightarrow (P_+
\mat{V}^{1/2 \dagger} \exp (-i\sum_A \alpha_A \mat{\lambda}_A)
\mat{V}^{1/2} + P_- \mat{V}^{1/2} \exp (-i\sum_A \alpha_A
\mat{\lambda}_A) \mat{V}^{1/2\dagger})\mat{\psi_j}$.

Using the same argument as above, we can apply a finite unitary
renormalization of fermion fields so that $\mat{M}_1$ and $\mat{M}_2$
are hermitian in each order of perturbation theory.  Assuming this has
been done, we can use (\ref{stat2}) to write,
\begin{equation}\label{pane}
  \mat{M}_{j\naught} = \mat{U}_{mjL}^\dagger (\mat{M}_j +
  \overline{\mat{\delta}}\mat{M}_j )
  \mat{U}_{mjR},
\end{equation}
so that in $\LL_\psi$
\begin{gather}\label{pane2}
  (P_+ \mat{M}_{j\naught} + P_- \mat{M}_{j\naught}^\dagger) =
  \overline{\mat{\cal U}}_{mj} (\mat{M}_j +
  \overline{\mat{\delta}}\mat{M}_j ) \mat{\cal U}_{mj},
  \quad \text{with}\\
  \mat{\cal U}_{mj} = P_+ \mat{U}_{mjR} + P_- \mat{U}_{mjL}, \quad
  \overline{\mat{\cal U}}_{mj} = \gamma^0 \mat{\cal U}_{mj}^\dagger
  \gamma^0, \quad \mat{\cal U}_{mj} \mat{\cal U}_{mj}^\dagger =
  \mat{1} = \mat{\cal U}^\dagger_{mj} \mat{\cal U}_{mj}. \notag
\end{gather}
In OS these expressions are always valid since $\mat{M}_j$ are real
diagonal to all orders, therefore hermitian.  In \ms we can use
(\ref{pane}) at one loop, because the tree-level $\mat{M}_j$ have
been chosen hermitian.  The resulting one-loop renormalized mass
matrices will not be hermitian.  As mentioned above, we can restore
hermitianity at one loop, at a given renormalization scale, by means
of a finite unitary renormalization of fermion fields.  This is
analogous to the choice of one-loop off-diagonal entries of the mass
matrix in \S \ref{sec:msbar}.

The bare fermion fields are given in terms of renormalized ones by
expressions similar to (\ref{psi}),
\begin{gather}
  \label{lab2}
  \mat{\psi}_{j\naught} = \mat{\cal U}_j \mat{\cal Z}_j^{1/2}
  \mat{\psi}_j \\
  \mat{\cal Z}_j^{1/2} = (P_+ \mat{Z}_{jR}^{1/2} + P_-
  \mat{Z}_{jL}^{1/2}), \quad
  \mat{\cal U}_j=\left( P_+ \mat{U}_{jR} + P_- \mat{U}_{jL}\right),
  \quad
  \mat{\cal U}_{j} \mat{\cal U}_{j}^\dagger = \mat{1}.
\end{gather}
The mixing matrix \mat{V} introduced in (\ref{lab1}) is renormalized
according to (\ref{stat}),
\begin{equation}\label{dow}
  \mat{V}_\naught= \mat{W}^\dagger \mat{Z}_V \mat{V}
  \mat{W}, 
  \quad
  \left[ \mat{Z}_V,\mat{V} \right]=0,
  \quad
  \mat{W} = e^{\displaystyle -i\mat{\delta W}}.
\end{equation}
Since $\mat{Z}_V$ is unitary, we can write $\mat{Z}_V =
\exp(-i\mat{\delta Z}_V)$, with $\mat{\delta Z}_V$ hermitian, $\left[
\mat{\delta Z}_V,\mat{V} \right]=0$.  Other parametrizations for the
mixing matrix counterterms are discussed in the next section.

Substituting these expressions for bare quantities in $\LL$, we obtain
its expression in terms of renormalized parameters and fields.  The
fermion and Yukawa Lagrangians, in particular, read
\begin{equation}\label{lab3}
  \begin{split}
    \LL_\psi+\LL_Y &= \sum_j \overline{\mat{\psi}}_j \overline{\mat{\cal
        Z}}_j^{1/2} i \dirac \mat{\cal Z}_j^{1/2} \mat{\psi}_j -
    \sum_j \overline{\mat{\psi}}_j \overline{\mat{\cal Z}}_j^{1/2}
    \overline{\mat{\cal U}}_j \overline{\mat{\cal U}}_{mj} (\mat{M} +
    \overline{\mat{\delta}} \mat{M}_j) \mat{\cal U}_{mj} \mat{\cal
      U}_j \mat{\cal Z}_j^{1/2} \mat{\psi}_j \\
    &\quad + \mu^{\epsilon/2} Z_g Z_\phi^{1/2} g
    \overline{\mat{\psi}}_1 P_+ \mat{Z}_{1L}^{1/2}
    \mat{U}_{1L}^\dagger \mat{W}^\dagger \mat{Z}_V \mat{VW}
    \mat{U}_{2R} \mat{Z}_{2R}^{1/2} \mat{\psi}_2 \phi +
    \text{H.c.}
  \end{split}
\end{equation}
Just as in \S \ref{sec:hermitean}, out of the unitary matrices
$\mat{\cal U}_j$, $\mat{\cal U}_{mj}$, $\mat{W}$ that we have
introduced, only the combinations $\mat{\cal U}_{mj} \mat{\cal U}_j$,
$\mat{W} \mat{U}_{1L}$ and $\mat{W} \mat{U}_{2R}$ enter $\LL$.
Clearly, we can always choose $\mat{\cal U}_j = 1$.  In OS, however,
it is convenient to set $\mat{\cal U}_{mj} = \mat{1}$ instead, so that
bare mass matrices are diagonal in the mass basis.

We see from (\ref{lab3}) that for the interaction term to retain its
form, with a unitary mixing matrix, we need $\mat{Z}_{1L},
\mat{Z}_{2R}\propto \mat{1}.$ This is the case in \ms, due to $SU(N)$
flavor symmetry.  In OS finite asymmetric counterterms to the Yukawa
coupling are needed, so the form of the Lagrangian is not preserved. 

\subsection{Other parametrizations for the mixing matrix}
\label{sec:otherpar}

The parametrization of counterterms given in (\ref{dow}) conforms to
the general form for normal matrices given in (\ref{stat}).  Since
$\mat{V}_\naught$ and \mat{V} are both unitary, however, we can write
the relation between them in other ways.  Clearly, $\mat{V}_\naught =
\mat{WV}$ or $\mat{V}_\naught = \mat{VW}$ are admissible since we can
reach any unitary matrix in a neighborhood of \mat{V} by varying
\mat{W} over a neighborhood of the identity in $SU(N)$.  Another usual
way of writing the renormalization constants for $\mat{V}$ is
$\mat{V}_\naught = \mat{W}_1 \mat{VW}_2$, with $\mat{W}_{1,2}$ unitary
\cite{sac,den}.  This parametrization is convenient from the
calculational point of view.  Here we point out that it is
overspecified, it must satisfy constraint relations analogous to those
considered in \S \ref{sec:overspec}, as we show next.

Given a map $\mat{F} : SU(N)\rightarrow SU(N)$ of the form
$\mat{F}(\mat{V}) = \mat{W}_1 \mat{V} \mat{W}_2$, $\mat{W}_j(\mat{V})
\in SU(N)$, we can always find $\widetilde{\mat{W}}_{1,2}$ such that,
\begin{equation}\label{alterp}
  \mat{F}(\mat{V}) = \widetilde{\mat{W}}_{1} \mat{V} \widetilde{\mat{W}}_{2},
  \quad
  \widetilde{\mat{W}}_{j} = \widetilde{\mat{W}}_{j} (\mat{V}) \in SU(N),
  \quad
  [\widetilde{\mat{W}}_{2}\widetilde{\mat{W}}_{1},\mat{V}] = 0 =
  [\widetilde{\mat{W}}_{1}\widetilde{\mat{W}}_{2},\mat{F}(\mat{V})].
\end{equation}
To see this we notice that, given $\mat{W}_{1,2}$, we can use Lemma
\ref{flem} of appendix \ref{sec:appb} to write $\mat{F}(\mat{V})$ as
in (\ref{dow}) (or (\ref{stat})).  With the same notation as in eq.\ 
(\ref{dow}), setting $\widetilde{\mat{W}}_1 = \mat{W}^\dagger
\mat{Z}_V$ and $\widetilde{\mat{W}}_2 = \mat{W}$ we get
(\ref{alterp}).

\subsection{On-shell scheme}
\label{OSunit}

We consider only the case of regular mass matrices $\mat{M}_j$.  The
extension to degenerate $\mat{M}_j$ can be carried out as in \S
\ref{sec:degenerate}.  Renormalization conditions in this scheme break
flavor symmetry, so counterterms are not symmetric, their finite parts
being tuned so the field basis is such that $\mat{M}_j$ are diagonal
to all orders.  Setting $\mat{\cal U}_{mj}=\mat{1}$ in (\ref{pane2})
we have $\mat{M}_{j\naught} = \mat{M}_j + \overline{\delta}\mat{M}_j$,
with $\overline{\delta}\mat{M}_j$ diagonal. To one-loop order we then
have, 
\begin{equation}\label{oslag}
  \begin{split}
    \LL_\psi &= \sum_j \overline{\mat{\psi}}_j i\dirac \mat{\psi}_j +
    \sum_j \overline{\mat{\psi}}_j i\dirac \left(P_+ \delta
      \mat{Z}_{jR}
      + P_- \delta \mat{Z}_{jL} \right) \mat{\psi}_j \\
    &\quad -\sum_j \overline{\mat{\psi}}_j \mat{M}_j \mat{\psi}_j -
    \sum_j \overline{\mat{\psi}}_j \left( P_+ \mat{\Delta M}_j + P_-
      \mat{\Delta M}_j^\dagger \right) \mat{\psi}_j\\
    \LL_Y &= \mu^{\epsilon/2} g \overline{\mat{\psi}}_1 P_+ \left(
    \mat{V} + \mat{\Delta V} + \mat{V} \Delta g +
    \mat{\Delta} \mat{\Gamma} \right) \mat{\psi}_2 \phi
  + \text{H.c.} \\
  \LL_\mathrm{em} &= Q_\phi^2 \mu^\epsilon e^2
  \left(1+\text{Re}(\delta Z_\phi)\right) 
    \phi^\dagger \phi A_\mu A^\mu - \mu^{\epsilon/2}
    j_\phi^\mu \left(1+\text{Re}(\delta Z_\phi)\right) A_\mu \\
    &\quad + \sum_j \mu^{\epsilon/2} j_j^\mu A_\mu + \sum_j Q_j
    \mu^{\epsilon/2} e \overline{\mat{\psi}}_j \gamma^\mu \left( P_-
    \mat{\delta Z}_{jL} + P_+ \mat{\delta Z}_{jR} \right) \mat{\psi}_j
    A_\mu,
  \end{split}
\end{equation}
where the counterterms are defined as,
\begin{equation}
  \label{eq:cttuni}  
  \begin{split}
    \mat{\Delta M}_j &= \overline{\mat{\delta}} \mat{M}_j + i
    \mat{\delta U}_{jL} \mat{M}_j - i \mat{M}_j \mat{\delta U}_{jR} +
    \frac{1}{2} \mat{\delta Z}_{jL} \mat{M}_j + \frac{1}{2} \mat{M}_j
    \mat{\delta Z}_{jR}\\
    \mat{\Delta V} &= i [\mat{\delta W},\mat{V}] - i \mat{\delta Z}_V
    \mat{V} + i \mat{\delta U}_{1L} \mat{V} - i \mat{V} \mat{\delta
      U}_{2R} \\
    \Delta g &= \delta Z_g + \frac{1}{2} \delta Z_\phi ~;
    \qquad
    \mat{\Delta} \mat{\Gamma} = \frac{1}{2} \mat{\delta
    Z}_{1L} \mat{V} + \frac{1}{2} \mat{V} \mat{\delta Z}_{2R}.
  \end{split}
\end{equation}
The flavor-asymmetric counterterms must be finite, divergent terms
being flavor symmetric.  In particular, the divergent parts of
wave-function renormalization $(\mat{\delta
  Z}_{jL,R})_\mathrm{div}\propto \mat{1}$.  The phase of $\delta
Z_\phi$ appears only in $\Delta g$, so it can be adjusted to keep
$\Delta g$ real.  We defined \mat{\Delta V} so that it is
perturbatively unitary.  In a flavor-symmetric, mass-independent
scheme such as \mbox{\ms,} fermion wave-function renormalization
constants are flavor scalars that can be absorbed in $\Delta g$, no
other counterterms to the Yukawa coupling being needed.  In
particular, there are no flavor-breaking dimension 4 operators in
$\LL$.  In OS, however, we also need $\mat{\Delta} \mat{\Gamma}$, whose
finite part can be viewed as either breaking the unitarity of \mat{V}
or the scalar nature of $g$.

For the computation of Feynman graphs involving fermion loops with
$\gamma_5$ vertices we use 't~Hooft and Veltman's prescription, with
$\gamma_5$ anticommuting with $\gamma^\mu$ for $\mu=0,\ldots,3$ and
commuting other\-wise.  We henceforth set $Q_\phi=0$ for the sake of
simplicity.  For the scalar field self-energy, with OS
renormalization conditions, we find,
\begin{align}
  \Pi_\phi(p^2) &= \Omega_\phi(p^2) - \Omega_\phi(m_\phi^2) -
  (p^2-m_\phi^2) \Omega_\phi^\prime(m_\phi^2) \\
  \text{Re}(\delta Z_\phi) &= -\frac{g^2 N}{16\pi^2}
  \left(\frac{2}{\epsilon} - \frac{1}{3}\right) +
  \Omega_\phi^\prime(m_\phi^2)  \notag\\
  \delta m_\phi^2 &= -\frac{g^2}{16\pi^2} \left(\frac{4}{\epsilon}
    \left( \sum_{ja}m_{ja}^2-\frac{N}{2} m_\phi^2\right)
  +\frac{N}{3} m_\phi^2 - \sum_{ja} m_{ja}^2 (1+a_0(m_{ja}^2))\right)
  + \Omega_\phi(m_\phi^2) \notag\\
    \Omega_\phi(p^2) &= \frac{g^2}{16\pi^2} \sum_{a,b} \mat{V}_{ab}
    \mat{V}_{ba}^\dagger (p^2-m_{1a}^2-m_{2b}^2)
    b_0(p^2,m_{1a}^2,m_{2b}^2).
    \notag
\end{align}
The fermion two-point function for each family can be expressed in
terms of form factors as,
\begin{equation}
  \begin{split}
  \mat{\Gamma}_j(p^2) &= \pirac -\mat{M}_j-\mat{\Pi}_j(p^2), \\
  \mat{\Pi}_j(p) &= (\pirac \mat{\Sigma}_j^L(p^2) + \mat{\Delta
      M}_j^\dagger ) P_- + (\pirac \mat{\Sigma}_j^R(p^2) +
    \mat{\Delta M}_j) P_+ + \mat{\Sigma}_j^S (p^2).
  \end{split}
\end{equation}
The form factors for the first family are given by,
\begin{equation}\label{unren}
  \begin{split}
    \Sigma_{1ab}^L (p^2) &= -\delta Z_{1ab}^L
    -\frac{g^2}{16\pi^2\epsilon} \delta_{ab} - Q_1^2 e^2
    \frac{1+\xi}{8\pi^2\epsilon} \delta_{ab} + \Omega_{1ab}^Y (p^2) +
    Q_1^2 \Omega_\mathrm{em}^V(p^2,m_\gamma^2,m_{1a}^2) \delta_{ab} \\
  \Sigma_{1ab}^R (p^2) &= -\delta Z_{1ab}^R - Q_1^2 e^2
  \frac{1+\xi}{8\pi^2\epsilon} \delta_{ab} +
  Q_1^2 \Omega_\mathrm{em}^V(p^2,m_\gamma^2,m_{1a}^2) \delta_{ab} \\
  \Sigma_{1ab}^S (p^2) &= m_{1a} Q_1^2 e^2
  \frac{3+\xi}{8\pi^2\epsilon} \delta_{ab} + m_{1a} Q_1^2
  \Omega_\mathrm{em}^S(p^2,m_\gamma^2,m_{1a}^2) \delta_{ab}.
  \end{split}
\end{equation}
It is not difficult to check that the dependence on $\xi$ in the
scalar and fermion propagators satisfies $U(1)_\mathrm{e.m.}$ Ward
identities to one-loop level \cite[\S 18.7]{zin}.  Notice the flavor
dependence of e.m.\ corrections in (\ref{unren}).  This flavor
structure is also apparent in the counterterms to the e.m.\ current in
(\ref{oslag}).  In equation (\ref{unren}) we have introduced the
quantities,
\begin{subequations}
  \begin{align}
    \Omega_{1ab}^Y(p^2) &= \frac{g^2}{16\pi^2} \sum_c V_{ac}
    V_{cb}^\dagger
    b_-(p^2,m_\phi^2, m_{2c}^2)\\
    \Omega_\mathrm{em}^V (p^2,m_\gamma^2,m_c^2) &= \frac{e^2}{16\pi^2}
    \left(\rule{0ex}{3ex} 1 + 2 b_-(p^2,m_\gamma^2,m_c^2)
      -b_0(p^2,m_\gamma^2,m_c^2) +
      \xi b_0 (p^2,\xi m_\gamma^2, m_c^2) \right.\\
    &\quad \left. + \frac{p^2-m_c^2}{m_\gamma^2} \left(
        b_1(p^2,m_\gamma^2,m_c^2) -
        b_1(p^2,\xi m_\gamma^2,m_c^2) \right) \right)\\
    \Omega_\mathrm{em}^S (p^2,m_\gamma^2,m_c^2) &= -\frac{e^2}{16\pi^2}
    \left( 2 + 3 b_0(p^2,m_\gamma^2,m_c^2) + \xi b_0(p^2,\xi
      m_\gamma^2,m_c^2) \right).\\
\intertext{The corresponding expressions for the second family are obtained by
changing the family index 1$\rightarrow$2 and L$\leftrightarrow$R,
with,}
  \Omega_{2ab}^Y(p^2) &= \frac{g^2}{16\pi^2} \sum_c V_{ac}^\dagger
    V_{cb} b_-(p^2,m_\phi^2, m_{1c}^2).
  \end{align}
\end{subequations}
The renormalization conditions are
\begin{subequations}
  \begin{gather}
  \Sigma_{jaa}^{L} (m_{ja}^2) + \Sigma_{jaa}^{R} (m_{ja}^2) + 2 m_{ja}
  \frac{\partial}{\partial p^2} \left[ m_{ja} \Sigma_{jaa}^{L}
  (p^2) + m_{ja} \Sigma_{jaa}^{R} (p^2) + 2 \Sigma_{jaa}^{S}
  (p^2) \rule{0ex}{3ex}\right]_{p^2=m_{ja}^2} = 0,
\intertext{for diagonal functions and}
  m_{jb} \Sigma_{jab}^{L,R} (m_{1b}^2) + \Delta M_{jab}^{R,L} +
  \Sigma_{jab}^{S} (m_{jb}^2) = 0,
  \qquad\
  m_{ja} \Sigma_{jab}^{L,R} (m_{1a}^2) + \Delta M_{jab}^{L,R} +
  \Sigma_{jab}^{S} (m_{ja}^2) = 0,
  \end{gather}  
\end{subequations}
for both diagonal and off-diagonal ones \cite{den}.  Here we have used
the notation $\Delta M_{jab}^{R} \equiv \Delta M_{jab}$ and $\Delta
M_{jab}^{L} \equiv \Delta M_{jab}^\dagger$ for brevity.  The
renormalization constants are then, in the flavor-diagonal case,
\begin{subequations}
  \begin{align}
    \delta Z_{1aa}^R &= -\frac{Q_1^2 e^2 \xi}{8\pi^2\epsilon} + Q_1^2
    \Omega_\mathrm{em}^V (m_{1a}^2,m_\gamma^2,m_{1a}^2) +
    m_{1a}^2 \Omega_{1aa}^{Y'} (m_{1a}^2) \notag\\
    &\quad + 2 m_{1a}^2 Q_1^2 \Omega_\mathrm{em}^{V'}
    (m_{1a}^2,m_\gamma^2,m_{1a}^2) + 2 m_{1a}^2
    Q_1^2 \Omega_\mathrm{em}^{S'} (m_{1a}^2,m_\gamma^2,m_{1a}^2)\\
    \delta Z_{1aa}^L &= -\frac{g^2}{16\pi^2\epsilon} + \Omega_{1aa}^Y
    (m_{1a}^2) + \delta Z_{1aa}^R \\
    \Delta M_{1aa} &= -m_{1a} Q_1^2 \Omega_\mathrm{em}^S
    (m_{1a}^2,m_\gamma^2,m_{1a}^2) -m_{1a} Q_1^2 \Omega_\mathrm{em}^V
    (m_{1a}^2,m_\gamma^2,m_{1a}^2) + m_{1a} \delta Z_{1aa}^R.
\intertext{In the off-diagonal case e.m.\ contributions vanish and
  we obtain the simpler expressions,}
    \delta Z_{1ab}^R &=
    \frac{m_{1a}m_{1b}}{m_{1a}^2-m_{1b}^2}
    \left(\Omega_{1ab}^Y(m_{1a}^2) - \Omega_{1ab}^Y(m_{1b}^2)\right) \\
    \delta Z_{1ab}^L &= \frac{m_{1a}^2 \Omega_{1ab}^Y(m_{1a}^2) -
      m_{1b}^2
      \Omega_{1ab}^Y(m_{1b}^2)}{m_{1a}^2-m_{1b}^2} \\
    \Delta M_{1ab} &= m_{1a} \delta Z_{1ab}^R.
  \end{align}  
\end{subequations}
We also obtain $\Delta M_{1ab}^L = m_{1b} \delta Z_{1ab}^R$, which is
consistent with the above definitions.  Notice that off-diagonal
counterterms are finite, as required by flavor symmetry.  Substituting
these results back into (\ref{unren}) we obtain the renormalized
fermion self-energy for the first family.  

In order to compute $\mat{\delta U}_1^{L,R}$ we proceed as in \S
\ref{sec:hermitean}.  It is convenient to split $\mat{\Delta M}$ into
its hermitian and anti-hermitian parts and, using (\ref{eq:cttuni}),
write an equation analogous to (\ref{fitt}) for each part.  Proceeding
in that way we get,
\begin{subequations}
  \begin{align}
    \overline{\delta} M_{1ab} &= \delta_{ab} \frac{m_{1a}}{2}
    \left(\delta
      Z_{1aa}^R - \delta Z_{1aa}^L \right) \\
    \delta U_{1ab}^R &= \frac{i}{2} \frac{m_{1a}^2 +
      m_{1b}^2}{m_{1a}^2-m_{1b}^2} \delta Z_{1ab}^R - i
    \frac{m_{1a}m_{1b}}{m_{1a}^2 - m_{1b}^2} \delta Z_{1ab}^L \\
    \delta U_{1ab}^L &= i \frac{m_{1a}m_{1b}}{m_{1a}^2
      - m_{1b}^2} \delta Z_{1ab}^R - \frac{i}{2} \frac{m_{1a}^2 +
      m_{1b}^2}{m_{1a}^2-m_{1b}^2} \delta Z_{1ab}^L,
  \end{align}
\end{subequations}
where in the last two lines $a\neq b$.  Diagonal elements of
$\mat{\delta U}_1^{L,R}$ are not determined by renormalization
conditions, and are set to vanish.  Renormalization constants for the
second family are given by similar expressions, after changing the
family index and exchanging $L$ and $R$ labels.  

We denote the three-point 1-PI function corresponding to the Yukawa
vertex as $\mat{\Gamma}_3(p_1,p_2)$.  External momenta are assumed to
be on their mass shell, $p_1^2=m_{1a}^2$, $p_2^2=m_{2b}^2$, and
$p_\phi^2=(p_1+p_2)^2=m_\phi^2$. $\mat{\Gamma}_3$ can be decomposed in
form factors as $\mat{\Gamma}_3=\mat{\Gamma}^+ P_+ + \mat{\Gamma}^-
P_- + \mat{\Gamma}_{3\mu}^V \gamma^\mu + \mat{\Gamma}_{3\mu}^A
\gamma^\mu\gamma_5 + \mat{\Gamma}_{3\mu\nu}^T \sigma^{\mu\nu}$.  Only
$\mat{\Gamma}_3^+$ receives divergent contributions at one loop,
\begin{equation}\label{form+}
  \Gamma_{3,ab}^+ = g (\mat{V} + \mat{V}
  \Delta g + \mat{\Delta V}+ \mat{\Delta} \mat{\Gamma})_{ab}
  + g e^2 Q^2 V_{ab} {\cal I}(p_1,p_2;m_{1a},m_{2b}),   
\end{equation}
where the loop integral ${\cal I}$ is defined in appendix
\ref{sec:appc}.  We separate in $\mat{\Delta} \mat{\Gamma}$ the  
contributions coming from flavor-diagonal fermion wave-function
renormalization constants, which are divergent and gauge-dependent,
from the finite, $\xi$-independent off-diagonal ones. 
Thus,
\begin{subequations}
  \begin{align}
  \Gamma_{3,ab}^+ &= g (\mat{V} + \mat{V}
  \Delta g + \mat{\Delta V})_{ab} + g
  \left(\widehat{\mat{\Delta}} \mat{\Gamma}\right)_{ab} 
  +g V_{ab} {\cal G}_{ab}\\
  \left(\widehat{\mat{\Delta}} \mat{\Gamma}\right)_{ab}
  &= \frac{1}{2} \sum_{c\neq a} \delta Z_{1ac}^L V_{cb} +
  \frac{1}{2} \sum_{c\neq b} V_{ac} \delta Z_{2cb}^R \\
  {\cal G}_{ab} &= \frac{1}{2} \delta Z_{1aa}^L + \frac{1}{2} \delta
  Z_{2bb}^R + e^2 Q^2 {\cal I}(p_1,p_2;m_{1a},m_{2b}).   
  \end{align}
\end{subequations}
An explicit expression for ${\cal G}_{ab}$ is as follows,
\begin{equation}\label{gab}
  \begin{split}
    {\cal G}_{ab} &= -\frac{g^2}{16\pi^2 \epsilon} + \frac{3 Q^2
      e^2}{8\pi^2 \epsilon} + \frac{1}{2}
    \left(\Omega_{1aa}^Y(m_{1a}^2) + m_{1a}^2
      \Omega_{1aa}^{Y'}(m_{1a}^2) + (m_{1a}^2\rightarrow
      m_{2b}^2) \right) \\
    &\quad +\frac{Q^2e^2}{16\pi^2}
    \left(-\frac{1}{2}\Lambda(m_{1a}^2,m_\gamma^2) -
      \frac{1}{2}\Lambda(m_{2b}^2,m_\gamma^2) +
      \Lambda_\xi(m_{1a}^2,m_\gamma^2) +
      \Lambda_\xi(m_{2b}^2,m_\gamma^2)\right)\\
    &\quad + 3 Q^2 e^2p_1\cdot p_2 C_0(p_\phi,p_1;
    m_\gamma^2,m_{1a}^2, m_{2b}^2) + Q^2 e^2 p_1\cdot p_2 \xi
    C_0(p_\phi,p_1; \xi m_\gamma^2,m_{1a}^2, m_{2b}^2),
  \end{split}
\end{equation}
where,
\begin{equation}
  \begin{split}
    \Lambda(m_{1a}^2,m_\gamma^2) &= 1 +
    b_1(m_{1a}^2,m_\gamma^2,m_{1a}^2) + 2
    b_0(m_{1a}^2,m_\gamma^2,m_{1a}^2) +4 m_{1a}^2 b_0'
    (m_{1a}^2,m_\gamma^2,m_{1a}^2) \\
    &\quad +4 m_{1a}^2 b_1' (m_{1a}^2,m_\gamma^2,m_{1a}^2),\\
    \Lambda_\xi(m_{1a}^2,m_\gamma^2) &= \frac{m_{1a}^2}{m_\gamma^2}
    \left( b_1(m_{1a}^2,m_\gamma^2,m_{1a}^2) -
    b_1(m_{1a}^2,\xi m_\gamma^2,m_{1a}^2)\right),
  \end{split}
\end{equation}
and $p_1\cdot p_2 = 1/2 (m_\phi^2-m_{1a}^2-m_{2b}^2)$.  Some comments
regarding ${\cal G}_{ab}$ are in order.  The ultraviolet divergent
terms in (\ref{gab}) are $\xi$-independent, as expected since they
must be cancelled by $\Delta g$.  All the remaining gauge dependence
in ${\cal G}_{ab}$ is contained in $\Lambda_\xi$ and $C_0$.  As is
well-known, and is clear from its definition in appendix
\ref{sec:appc}, the triangle integral $C_0$ with on-shell external
momenta diverges logarithmically as $m_\gamma^2\rightarrow 0$.  So
does $\Lambda_\xi$ as well, since we can write,
\begin{equation*}
  \Lambda_\xi(m^2,m_\gamma^2) = \frac{1}{2}
  \left(b_0(m^2,m_\gamma^2,m^2) - a_0(m_\gamma^2)\right) -
  \frac{\xi}{2} \left(b_0(m^2,\xi m_\gamma^2,m^2) - a_0(\xi
    m_\gamma^2)\right), 
\end{equation*}
with
\begin{equation*}
  b_0(m^2,m_\gamma^2,m^2) - a_0(m_\gamma^2) =
  \int_0^1 \! dx\, \log\left(1-x+\frac{m^2}{m_\gamma^2}x^2\right) + 1. 
\end{equation*}
These infrared divergences must be cancelled in the computation of
transition rates by the contribution from real-photon emission
diagrams.  We conclude that the infrared divergent terms must appear
explicitly in the amplitude, and should not be absorbed in the finite
part of counterterms.  In fact, the requirements of infrared and
ultraviolet finiteness, that $g$ and $g_\naught$ should be
flavor-independent and $\mat{V}$ and $\mat{V_0}$ unitary, and of gauge
invariance, lead us to set,
\begin{equation}
 \Delta g = - ({\cal G}_{ab})_\mathrm{div} 
          = \frac{g^2}{16 \pi^2 \epsilon} - \frac{3 Q^2 e^2}{8\pi^2
          \epsilon},  
 \qquad\text{and}\qquad
 \mat{\Delta V} = 0.
\end{equation}
These equations determine the renormalization constants $\delta Z_g$,
$\mat{\delta W}$ and $\mat{\delta Z}_V$ through (\ref{eq:cttuni}) and
the above results for wave-function renormalization constants.  Our
choice is not unique, though, since the above requirements still allow
the finite part of $\delta Z_g$ to be redefined by adding gauge- and
flavor-independent arbitrary constants.  We could, for instance,
define $g$ through the total $\phi$ decay width, or some other
inclusive process.  We will not pursue such phenomenological analysis
further here.

\subsection{\ms scheme}

Renormalization in \ms scheme turns out, as expected, to be much
simpler than in OS.  The renormalized Lagrangian at one-loop level
takes essentially the same form as in (\ref{oslag}), (\ref{eq:cttuni}).
The main differences being that now $\mat{\delta U}_j^{L,R} =
\mat{0}$, $\mat{\delta Z}_j^{L,R} = \delta Z_j^{L,R} \mat{1}$, and
$\mat{\delta U}_{mj}^{L,R}\neq \mat{0}$.  The mass matrices
$\mat{M}_j$ are diagonal at tree level but not necessarily diagonal,
or even hermitian, at one loop.  We are then led to separate diagonal
elements (which are real, given our choice of tree-level flavor basis)
from off-diagonal ones (which are ${\cal O}(g^2)$ or ${\cal O}(e^2)$),
as discussed in \S \ref{sec:msbar}. 

We define $\Delta g=\delta Z_g + 1/2 \delta Z_\phi + 1/2 \delta Z_1^L
+ 1/2 \delta Z_2^R$, and set $\mat{\Delta} \mat{\Gamma} =
\mat{0}$, since obviously no asymmetric counterterms are needed in
this scheme.  Furthermore, $\mat{\delta W}=0$ for the same reasons as
in \S \ref{sec:msbar}.  Renormalization constants can then be read off
the corresponding expressions in \S \ref{OSunit}.  In particular,
\begin{equation}
  \delta Z_1^L = -\frac{Q^2 e^2 \xi}{8 \pi^2 \epsilon} - \frac{g^2}{16
    \pi^2 \epsilon} = \delta Z_2^R,
  \qquad
  \delta Z_g = \frac{g^2 (N+1)}{16 \pi^2 \epsilon} - \frac{3 Q^2
    e^2}{8 \pi^2 \epsilon}.
\end{equation}
Divergent contributions from the triangle diagram correcting the
Yukawa vertex, which are flavor symmetric, have been absorbed in
$\delta Z_g$.  There are no further infinities to cancel, so that
$\mat{\delta Z}_V=\mat{0}$, \emph{i.e.,} \mat{V} is not renormalized.

We can easily derive renormalization group equations for renormalized
parameters and Green's functions.  However, \mat{V} obviously does not
run at one loop in this model.
\section{Concluding Remarks}
\label{sec:final}

Unitary transformations play a distinguished role in quantum theories,
as is well known.  In QFT, unitary transformations in some internal
``flavor'' space preserve the normalization of kinetic terms and
currents, in particular electromagnetic ones.  It is then natural to
consider the class of mixing matrices that can be diagonalized by a
unitary transformation, namely, normal mixing matrices, and their
renormalization properties.  This is the subject of the foregoing.

We point out in \S \ref{sec:hermitean} and \S \ref{sec:unitary} that
the usual parametrizations for mixing matrix counterterms
\cite{sac,kni,gam,bar} are overspecified.  This fact, of course, does
not make them less useful in any way, but had not been pointed out in
the previous literature.  We explicitly exhibit the constraints those
parametrizations must satisfy, and their solution.  Furthermore, the
minimal parametrization given in appendix \ref{sec:appb} can be
applied also to the case of normal mixing matrices, which we do in \S
\ref{sec:normal}.  

In principle, the general case of non-singular, not necessarily normal
mixing matrices, although not discussed in this paper, can also be
studied in the framework proposed here for normal matrices.  This is
so because any complex matrix \mat{A} can be decomposed as \mat{A} =
\mat{N} + \mat{T}, with \mat{N} normal and \mat{T} nilpotent such that
they can be simultaneously brought to diagonal and strictly triangular
form, respectively, by a unitary transformation.

Writing field-strength renormalization matrices in polar components is
useful in OS scheme, as discussed in \S 2 and \S 4.  At one-loop level
this is the same as decomposing those matrices into their hermitian
and anti-hermitian parts, as done in \cite{sac}.  As remarked in \S 2,
the unitary and hermitian components play different roles in the
theory.  Furthermore, the contribution to higher orders in
perturbation theory from powers of the one-loop \mat{\delta U} is
apparent in this way due to the exponential form of the unitary matrix
$\mat{U}=\exp(i \mat{\delta U})$.

\subsection*{Acknowledgements}

Discussions with L.\ Chan Palomo are gratefully acknowledged.  I would
like to thank C.\ Garc\'{\i}a Canal, V.\ Gupta and A.\ Zepeda for
reading the manuscript and for their helpful comments.  This work was
partially supported by SNI and by Conacyt Research Project 32598-E.

\appendix
\numberwithin{equation}{section}

\section{Mappings of normal matrices}
\label{sec:appb}

A matrix $\mat{N}\in \mathbb{C}^{N\times N}$ is called normal if
$[\mat{N}, \mat{N}^\dagger]=0$ \cite{hal}. It can be shown \cite{hal}
that a complex matrix can be diagonalized by a unitary transformation
if and only if it is normal.  We denote by $\mat{n}(N)$ the set of
$N\times N$ normal matrices. The Lie algebra $\mat{u}(N)$ of hermitian
matrices, the group $U(N)$ of unitary matrices, and the Abelian
algebra of diagonal matrices are all contained in $\mat{n}(N)$.

\begin{lem}\label{flem}
  Let $\mat{F}: \mat{n}(N)\rightarrow \mat{n}(N)$ be a mapping of
  normal matrices such that $\mathrm{rank}(\mat{F}(\mat{X})) \leq
  \mathrm{rank}(\mat{X})$ for every \mat{X} in $\mat{n}(N)$.  Then, we
  can write,
  \begin{equation}\label{stat}
    \mat{F}(\mat{X})=\mat{U}^\dagger\mat{Z}\mat{X}\mat{U},\quad
    \text{with}\quad
    [\mat{Z},\mat{X}]=0, \quad
    \mat{Z}=\mat{Z}(\mat{X}) \in \mat{n}(N),
    \quad \mat{U}=\mat{U}(\mat{X}) \in U(N).
  \end{equation}
\end{lem}
\begin{proof}
  We first consider the case of \mat{X} non-singular.  In fact, in
  this case there is an open neighborhood of \mat{X} in $\mat{n}(N)$
  where all matrices are non-singular.  We can apply the
  parametrization (\ref{stat}) to every matrix in that neighborhood.
  Using the normality of \mat{F}(\mat{X}) and \mat{X}, we can write
  (dropping the argument \mat{X} for brevity) $\mat{F} =
  \mat{V}^\dagger \mat{F}'\mat{V}$ and $\mat{X} = \mat{W}^\dagger
  \mat{X}'\mat{W}$, with $\mat{V}$, $\mat{W}$ unitary and $\mat{F}'$,
  $\mat{X}'$ diagonal.  Define $\mat{Z}'$ diagonal by
  $\mat{F}'=\mat{Z}'\mat{X}'$. Then $\mat{U}=\mat{W}^\dagger\mat{V}\in
  U(N)$ and $\mat{Z}=\mat{W}^\dagger\mat{Z}'\mat{W} \in \mat{n}(N)$
  satisfy (\ref{stat}).
  
  Assume now \mat{X} has a null eigenvalue with multiplicity $r$.  We
  can choose \mat{W} so that $\mat{X}' =
  \mathrm{diag}(0,\ldots,0,x_{r+1},\ldots,x_n)$ with $x_j \neq 0$.
  Similarly, we can choose \mat{V} so that $\mat{F}' =
  \mathrm{diag}(0, \ldots, 0, f_{q+1},$  $\ldots, f_n)$, with $f_j\neq 0$
  and $q\geq r$.  We then define $\mat{Z}' = \mathrm{diag}
  (1,\ldots,1, f_{r+1}/x_{r+1},\ldots,f_{n}/x_{n})$ and proceed as
  above.
\end{proof}

A transformation \mat{F} satisfying the rank hypothesis of the
Lemma may be called ``multiplicative,'' in the sense that $\mat{F}
(\mat{0}) = \mat{0}$.  The same argument as in the multiplicative
case, with obvious changes, proves the following.
\begin{lem}
  Let $\mat{F}: \mat{n}(N)\rightarrow \mat{n}(N)$ be a mapping of
  normal matrices.  Then, for any $\mat{X}\in \mat{n}(N)$ we can
  write,
  \begin{equation}\label{stat1}
    \mat{F}(\mat{X})=\mat{U}^\dagger(\mat{X}+\mat{N})\mat{U},\quad
    \text{with}\,
    [\mat{N},\mat{X}]=0, \quad
    \mat{N}=\mat{N}(\mat{X}) \in \mat{n}(N),
    \, \mat{U}=\mat{U}(\mat{X}) \in U(N).
  \end{equation}
\end{lem}

We remark that both (\ref{stat}) and (\ref{stat1}) hold for mappings
of hermitian matrices, $\mat{F}: \mat{u}(N)\rightarrow \mat{u}(N)$,
with \mat{Z} and \mat{N} hermitian, and (\ref{stat}) also holds for
mappings of unitary matrices, $\mat{F}: U(N)\rightarrow U(N)$, with
\mat{Z} unitary.

Consider, finally, a general matrix function of a hermitian matrix
$\mat{F} : \mat{u}(N) \rightarrow \mathbb{C}^{N\times N}$.  Using the
polar decomposition $\mat{F}(\mat{X}) = \mat{R}(\mat{X})
\mat{U}_F(\mat{X})$ with $\mat{U}_F$ unitary and \mat{R} hermitian,
and applying eqs.\ (\ref{stat}) and (\ref{stat1}) to \mat{R}, we
obtain the parametrizations,
\begin{subequations}  \label{stat2}  
  \begin{align}
  \mat{F}(\mat{X}) &= \mat{U Z X V}, \quad \mat{U}=\mat{U}(\mat{X} )
  \text{ and } \mat{V}=\mat{V}(\mat{X}) \in U(N), \;
  \mat{Z}=\mat{Z}(\mat{X}) \in \mat{u}(N), \;
  [\mat{Z},\mat{X}] = 0.\\
  \mat{F}(\mat{X}) &= \mat{U} (\mat{X} + \mat{\delta X}) \mat{V},
  \quad \mat{\delta X}=\mat{\delta X}(\mat{X} ) \in \mat{u}(N), \;
  [\mat{\delta X},\mat{X}] = 0.
  \end{align}
\end{subequations}
Substituting (\ref{stat2}) into the hermitian matrices
$\mat{FF}^\dagger$ and $\mat{F}^\dagger \mat{F}$, we recover
(\ref{stat}) and (\ref{stat1}).  

\section{Perturbative factorization of $SU(N)$ matrices}
\label{sec:appa}

In \S \ref{sec:hermitean} we make use of the fact that a unitary
matrix \mat{U} which is close to the identity can always be
factorized as $\mat{U} = \exp(i \mat{\delta U}') \exp(i
\mat{\delta \widetilde{U}})$, where $\mat{\delta U}'$ is hermitian and
diagonal and $\mat{\delta \widetilde{U}}$ is hermitian and has zeros
on the diagonal.  A sketch of the proof follows (see \cite{wei} for a
globally valid factorization). 

We decompose the Lie algebra of hermitian matrices as $\mat{u}(N) =
\mat{a} \oplus \mat{b}$, where \mat{a} is the Cartan subalgebra of
diagonal matrices of $\mat{u}(N)$ and \mat{b} its complementary
subspace.  Let $\mat{U} = \exp(i \epsilon \mat{H})$, with $\epsilon$
small and $\mat{H} \in \mat{u}(N)$.  We want to show that we can
always find  $\mat{A}=\mat{A}(\epsilon) \in \mat{a}$ and
$\mat{B}=\mat{B}(\epsilon) \in \mat{b}$ such that
\begin{equation}
  e^{i \epsilon \mat{A}} e^{i \epsilon \mat{B}} =
  \sum_{n=0}^N \frac{(i\epsilon)^n}{n!} \mat{H}^n + {\cal
  O}(\epsilon^{N+1}). 
\end{equation}
To this end, we write $\mat{A} = \sum_{n=0}^N \frac{\epsilon^n}{n!}
\mat{A}_n$ with $\mat{A}_n \in \mat{a}$, and analogously for \mat{B},
and consider the equation $\exp(i\epsilon\mat{H}) =
\exp(i\epsilon\mat{A}) \exp(i\epsilon\mat{B})$ which, using the
Campbell-Baker-Hausdorff formula (see \cite{jac,wei} and references
therein) can be written as,
\begin{equation}\label{one}
  \mat{H} = (\mat{A}+\mat{B}) + \frac{i\epsilon}{2} [\mat{A},\mat{B}]
  + \frac{(i\epsilon)^2}{12} \left([[\mat{A},\mat{B}],\mat{B}] +
    [[\mat{B},\mat{A}],\mat{A}]\right) + \cdots
\end{equation}
Expanding the rhs in powers of $\epsilon$, we are led to a set of
recursive equations,
\begin{equation}\label{two}
  \begin{split}
  \mat{A}_0 + \mat{B}_0 &= \mat{H}\\
  \mat{A}_1 + \mat{B}_1 &= \frac{1}{2} [\mat{B}_0,\mat{A}_0]\\
  \mat{A}_2 + \mat{B}_2 &= [\mat{B}_1,\mat{A}_0] +
  [\mat{B}_0,\mat{A}_1] - \frac{1}{6}
  [[\mat{A}_0,\mat{B}_0],\mat{B}_0] - \frac{1}{6}
  [[\mat{B}_0,\mat{A}_0],\mat{A}_0],~~\mathrm{etc.} 
  \end{split}
\end{equation}
which can be solved iteratively up to the desired order.  (Notice that
in (\ref{two}) we have used the Abelianity of \mat{a} to eliminate
some commutators.) Thus,
$\mat{A}_0 = \mathrm{diag}(H_{11},\ldots,H_{NN})$ and $\mat{B}_0 =
\mat{H} - \mat{A}_0$, $\mat{A}_1 = \mat{0}$ and $\mat{B}_1 = 1/2
[\mat{B}_0, \mat{A}_0]$, and so on.  More generally, the equation for
the coefficients of order $N+1$ obtained from (\ref{one}) is of the
form, 
\begin{equation*}
  \begin{split}
  0 &= \frac{1}{(N+1)!} (\mat{A}_{N+1}+\mat{B}_{N+1}) + \frac{1}{2}
  \sum_{\raisebox{0pt}[9pt][5pt]{$\begin{array}{c}
      \scriptstyle m,n=0\\[-5pt]
      \scriptstyle m+n=N\end{array}$}}^N  \frac{1}{m! n!}
  [\mat{A}_m,\mat{B}_n]\\ 
  &\quad + 
  \frac{1}{12}   \sum_{\raisebox{0pt}[9pt][5pt]{$\begin{array}{c}
      \scriptstyle k,m,n=0\\[-5pt]
      \scriptstyle k+m+n=N-1\end{array}$}}^{N-1}  \frac{1}{k! m!
  n!}\left( [[\mat{A}_k,\mat{B}_m], \mat{B}_n] -
  [[\mat{A}_k,\mat{B}_m],\mat{A}_n] \right) + \cdots 
  \end{split}
\end{equation*}
where the ellipsis refers to higher-order commutators. Terms with
$K$-order commutators involve $\mat{A}_j$, $\mat{B}_j$ with $j =
0,\ldots, N+1-K$, which are already known.  Thus, by projecting
$(\mat{A}_{N+1} + \mat{B}_{N+1})$ as given by this equation over
\mat{a} and \mat{b} we find $\mat{A}_{N+1}$ and $\mat{B}_{N+1}$.  

To summarize, let $D$ be a neighborhood of the identity in $U(N)$,
determining a corresponding neighborhood $\mat{d}$ of \mat{0} in
$\mat{u}(N)$, where the CBH formula holds.  Then, the elements of the
form $e^{\mat{A}} e^{\mat{B}}$ with $\mat{A}\in \mat{a}$ and
$\mat{B}\in \mat{b}$ are dense in $D$.
  
\section{Kernel and image of the adjoint map of a normal matrix}
\label{sec:kernel}

Consider a fixed $N\times N$ normal matrix $\mat{N}$.  Then, any
matrix $\mat{A}\in \mathbb{C}^{N\times N}$ can be written as
\begin{equation}\label{deco}
  \mat{A} = \mat{B} + [\mat{N},\mat{D}] \quad \text{with} \quad
  [\mat{N},\mat{B}] = 0.
\end{equation}
Equations (\ref{fitt}) and (\ref{fith}) are of this type.  Given
\mat{A} and \mat{N}, (\ref{deco}) can always be solved for \mat{B} and
\mat{D}, as we now show (see also \cite[\S 3.3]{jac}).

Since $\mat{N}$ is normal, it can be diagonalized by a unitary matrix.
Let $\lambda_1,\ldots, \lambda_r$ ($r\leq N$) be its eigenvalues and
$d_j (j=1,\ldots,r)$ their multiplicities.  The associated orthonormal
basis of $\mathbb{C}^N$ of eigenvectors of $\mat{N}$ is denoted by
$\{|\lambda_j, \alpha_j\rangle\}$ ($j=1,\ldots,r$,
$\alpha_j=1,\ldots,d_j$). Then we have $\mat{N}=\sum_{i\alpha}
\lambda_i |\lambda_i,\alpha\rangle \langle \lambda_i,\alpha|$, and for
any $\mat{A}\in \mathbb{C}^{N\times 
  N}$,
\begin{equation}
  \label{anya}
\mat{A} = \sum_{i,\alpha,j,\beta} A^{i\alpha}_{j\beta}
|\lambda_i,\alpha\rangle \langle \lambda_j,\beta|.
\end{equation}
We define the adjoint map associated to $\mat{N}$ as 
$\mathrm{ad}_N : \mathbb{C}^{N\times N} \rightarrow
\mathbb{C}^{N\times N}$,  
$\mathrm{ad}_N(\mat{A}) = [\mat{N},\mat{A}]$.  Then the set of
orthogonal projectors $\{|\lambda_i,\alpha_i\rangle \langle
  \lambda_j,\beta_j|\}$ is a basis of $\mathbb{C}^{N\times N}$ of
eigenvectors of $\mathrm{ad}_N$, since,
\(
\left[\mat{N},|\lambda_i,\alpha_i\rangle
  \langle\lambda_j,\beta_j|\right] = 
(\lambda_i -\lambda_j) |\lambda_i,\alpha_i\rangle
\langle\lambda_j,\beta_j| .
\) 
Therefore, we can rewrite (\ref{anya}) as
\begin{equation}\label{lst}
  \mat{A}  = 
  \sum_{i,\alpha,\beta}A^{i\alpha}_{i\beta}
  |\lambda_i,\alpha\rangle \langle \lambda_i,\beta| +
  \left[\mat{N},
    \sum_{\raisebox{0pt}[9pt][5pt]{$\begin{array}{c}
      \scriptstyle i,\alpha,j,\beta\\[-5pt]
      \scriptstyle i\neq j \end{array}$}}
    A^{i\alpha}_{j\beta}\frac{1}{\lambda_i-\lambda_j}
      |\lambda_i,\alpha\rangle \langle \lambda_j,\beta|\right],
\end{equation}
where the first term on the rhs obviously commutes with
$\mat{N}$.  This is the decomposition (\ref{deco}).

Whereas (\ref{lst}) provides an explicit solution to (\ref{deco}), a
slightly broader point of view can be adopted.  Consider the inner
product $(\mat{A},\mat{B})=\tr(\mat{A}^\dagger \mat{B})$ in
$\mathbb{C}^{N\times N}$.  By the cyclic property of the trace,
$(\mat{A},\mathrm{ad}_N (\mat{B})) =
(\mat{A},[\mat{N},\mat{B}]) =
\tr(\mat{A}^\dagger [\mat{N},\mat{B}]) =
\tr([\mat{N}^\dagger,\mat{A}]^\dagger \mat{B}) = 
([\mat{N}^\dagger,\mat{A}],\mat{B}) =
(\mathrm{ad}_{N^\dagger} (\mat{A}),\mat{B})$, so
$(\mathrm{ad}_N)^\dagger = \mathrm{ad}_{N^\dagger}$.  With this, using
the Jacobi identity and the normality of \mat{N}, it is immediate that
$[(\mathrm{ad}_N)^\dagger, \mathrm{ad}_N]=0$ and thus $\mathrm{ad}_N$
is a normal transformation of $\mathbb{C}^{N\times N}$. Therefore the
spectral theorem \cite{hal} holds for $\mathrm{ad}_N$.  In particular,
$\mathbb{C}^{N\times N} = \mathrm{Ker}(\mathrm{ad}_N) \oplus
\mathrm{Im}(\mathrm{ad}_N)$. This orthogonal decomposition shows that
solutions to (\ref{deco}) exist and are unique up to addition to
\mat{D} of a matrix commuting with \mat{N}.

\section{Loop integrals}
\label{sec:appc}

In this appendix we give a list of loop integrals used in the
foregoing.  More complete calculations can be found, e.g., in
\cite{den,vel,pve,ste}.
Divergent integrals are separated in a dimensional-regularization pole
term and a finite remainder.  $\mub = \mu\sqrt{4\pi e^{-\gamma_E}}$.

\begin{align*}
  A_0(m^2) &= \frac{i\mu^\epsilon}{(2\pi)^d} \int d^d\ell
  \frac{1}{\ell^2-m^2+i\varepsilon} = -\frac{m^2}{8\pi^2\epsilon} +
  \frac{m^2}{16\pi^2} a_0(m^2)\\
  a_0(m^2) &= \log\left(\frac{m^2}{\mub^2}\right) - 1\\
  A_1^\mu(m^2) &= \frac{i\mu^\epsilon}{(2\pi)^d} \int d^d\ell
  \frac{\ell^\mu}{\ell^2-m^2+i\varepsilon} = 0\\
  A_2^{\mu\nu}(m^2) &= \frac{i\mu^\epsilon}{(2\pi)^d} \int d^d\ell
  \frac{\ell^\mu\ell^\nu}{\ell^2-m^2+i\varepsilon} =
  -\frac{m^4}{32\pi^2\epsilon} g^{\mu\nu} + \frac{m^4}{64\pi^2}
  g^{\mu\nu} a_2(m^2)\\
  a_2(m^2) &= a_0(m^2) - \frac{1}{2}\\
  A_2(m^2) &= \frac{i\mu^\epsilon}{(2\pi)^d} \int d^d\ell
  \frac{\ell^2}{\ell^2-m^2+i\varepsilon} = m^2 A_0(m^2)\\
  B_0(p^\mu,m_1^2,m_2^2) &= \frac{i\mu^\epsilon}{(2\pi)^d} \int d^d\ell
  \frac{1}{\left(\ell^2-m_1^2+i\varepsilon\right)
    \left((\ell+p)^2-m_2^2+i \varepsilon\right)}\\
  &= -\frac{1}{8\pi^2\epsilon} + \frac{1}{16\pi^2}
  b_0(p^2,m_1^2,m_2^2)\\
  b_0(p^2,m_1^2,m_2^2) &= \int_0^1 dx \log\left((1-x)
  \frac{m_1^2}{\mub^2} + x \frac{m_2^2}{\mub^2} - x(1-x)
  \frac{p^2}{\mub^2} -i\varepsilon\right)\\
  B_1^\mu(p^\mu,m_1^2,m_2^2) &= \frac{i\mu^\epsilon}{(2\pi)^d} \int d^d\ell
  \frac{\ell^\mu}{\left(\ell^2-m_1^2+i\varepsilon\right)
    \left((\ell+p)^2-m_2^2+i \varepsilon\right)}\\
  &= \frac{p^\mu}{16\pi^2\epsilon} - \frac{p^\mu}{16\pi^2}
  b_1(p^2,m_1^2,m_2^2)\\
  b_1(p^2,m_1^2,m_2^2) &= \int_0^1 dx x \log\left((1-x)
  \frac{m_1^2}{\mub^2} + x \frac{m_2^2}{\mub^2} - x(1-x)
  \frac{p^2}{\mub^2} -i\varepsilon\right)\\
\intertext{which can also be written,}
  B_1^\mu(p^\mu,m_1^2,m_2^2) &= p^\mu B_1(p^2,m_1^2,m_2^2)\\
  p^2 B_1(p^2,m_1^2,m_2^2) &= \frac{1}{2}\left(A_0(m_1^2) - A_0(m_2^2)
    - (p^2+m_1^2-m_2^2) B_0(p^2,m_1^2,m_2^2)\right)\\
\intertext{We also use the combination $b_-(p^2,m_1^2,m_2^2)=b_0(p^2,m_1^2,m_2^2)
-b_1(p^2,m_1^2,m_2^2)$.}
  C_0(p_1,p_2,m_1^2,m_2^2,m_3^2) &=
  \frac{i\mu^\epsilon}{(2\pi)^d} \int d^d\ell
  \frac{1}{\left(\ell^2-m_1^2+i\varepsilon\right)
  \left((\ell+p_2)^2-m_2^2+i \varepsilon\right)
  \left((\ell-p_1+p_2)^2-m_3^2+i \varepsilon\right)}\\
\intertext{$C_0$ is ultraviolet finite.  In \S \ref{sec:OS} we use the
  following triangle integrals,}
 H_1(p_1,p_2;m_1^2,m_2^2,m_3^2) &= \frac{i\mu^\epsilon}{(2\pi)^d} \int
 d^d\ell \frac{1}{\ell^2-m_1^2+i\varepsilon} 
 \frac{\lirac+\pirac_2+m_2}{(\ell+p_2)^2-m_2^2+i\varepsilon}
 \frac{\lirac-\pirac_1+\pirac_2+m_3}{(\ell-p_1+p_2)^2-m_3^2+i\varepsilon}\\
 \tr(H_1(p_1,p_2;m_1^2,m_2^2,m_3^2)) &= \tr(1) \left(
   -\frac{1}{8\pi^2\epsilon} + h_1(p_1,p_2;m_1^2,m_2^2,m_3^2)\right)\\
 h_1(p_1,p_2;m_1^2,m_2^2,m_3^2) &=
 \frac{1}{2} b_0\left((p_1-p_2)^2;m_1^2,m_3^2\right) +
 \frac{1}{2} b_0\left(p_2^2;m_1^2,m_2^2\right) \\*
 &\quad +
 \frac{1}{2} \left((m_2+m_3)^2-p_1^2\right)
 C_0\left(p_1,p_2;m_1^2,m_2^2,m_3^2\right)\\
  h_2(p_1,p_2;m_1^2,m_2^2) &= \tr(1) m_2
  C_0\left(p_1,p_2;m_2^2,m_1^2,m_1^2\right) \\
\intertext{with $\tr(1)=4$. Finally, in \S \ref{OSunit} we define,}
  {\cal I}(p_1,p_2;m_{1a}^2,m_{2b}^2,m_\gamma^2) &=
  \frac{1}{2}\frac{\mu^\epsilon}{(2\pi)^d} \int d^d\ell
  \Delta_{\mu\nu}(\ell) 
  \frac{\tr\left(P_+ \gamma^\mu (\lirac+\pirac_1+m_{1a}) P_+
      (\lirac-\pirac_2+m_{2b}) \gamma^\nu \right)}
  {\left((\ell+p_1)^2-m_{1a}^2+i\varepsilon\right)
    \left((\ell-p_2)^2-m_{2b}^2+i\varepsilon\right)}\\
  = \frac{3+\xi}{8\pi^2\epsilon} - \frac{1}{8\pi^2} &-
  \frac{1}{32\pi^2}
  \left( 3 b_0(p_1^2,m_\gamma^2,m_{1a}^2) + 3
    b_0(p_2^2,m_\gamma^2,m_{2b}^2) + \xi
  b_0(p_1^2,\xi m_\gamma^2,m_{1a}^2) \right.\\
  \left. +\xi b_0(p_2^2,\xi m_\gamma^2, m_{2b}^2) \right) &
  + p_1\cdot p_2 \left(3 C_0(p_\phi, p_1; m_\gamma^2, m_{1a}^2,
  m_{2b}^2) + \xi C_0(p_\phi, p_1; \xi m_\gamma^2, m_{1a}^2,
  m_{2b}^2)\right) 
\end{align*}

\begin{thebibliography}{99}
\bibitem{sac} A.Denner, T.Sack, Nucl.Phys.\ \textbf{B347}, (1990),
  203.
%
\bibitem{kni} B.Kniehl, A.Pilaftsis, Nucl.Phys.\ \textbf{B474},
  (1996), 286.
%
\bibitem{gia} C.G.Bollini, J.J.Giambiagi, Phys.Lett.\  \textbf{B40},
  (1972), 566;  Nuov.Cim.\  \textbf{B12}, (1972), 270. 
%
\bibitem{den} A.Denner, Fort.Phys.\ \textbf{41}, (1993), 307.
%
\bibitem{gam} P.Gambino, P.Grassi, F.Madricardo, Phys.Lett.\ 
  \textbf{B454}, (1999), 98.
%
\bibitem{bar} A.Barroso, L.Br\"ucher, R.Santos,  Phys.Rev.\
  \textbf{D62}, (2000), 096003. 
%
\bibitem{col1} J.Collins, Nucl.Phys.\ \textbf{B80}, (1974), 341.
%
\bibitem{col2} J.Collins, ``Renormalization,'' Cambridge U.P.,
  Cambridge, 1992.
%
\bibitem{jac} N.Jacobson, ``Lie Algebras,'' Dover, New York, 1962.
%
\bibitem{cole} S.Coleman, ``Aspects of Symmetry,'' Cambridge U.P., New
  York, 1993.
%
\bibitem{zin} J.Zinn-Justin, ``Quantum Field Theory and Critical
  Phenomena,'' Oxford U.P., Oxford, 1997.
%
\bibitem{swe} S.Weinberg, ``The Quantum Theory of Fields,'' Vol.\ II,
  Cambridge U.P., New York, 1996.
%
\bibitem{hal} P.Halmos, ``Finite-Dimensional Vector Spaces,''
  Springer, New York, 1987.
%
\bibitem{wei} J.Wei, J.Math.Phys.\ \textbf{4}, (1963), 1337.
%
\bibitem{vel} G.'t Hooft, M.Veltman, Nucl.Phys.\ \textbf{B153},
  (1979), 365.
%
\bibitem{pve} G.Passarino, M.Veltman, Nucl.Phys.\ \textbf{B160},
  (1979), 151.
%
\bibitem{ste} R.Harlander, M.Steinhauser, Prog.Part.Nucl.Phys.\
  \textbf{43}, (1999), 167.
\end{thebibliography}
\end{document}


\maketitle
In a recent article \cite{bou} we discussed the renormalization of
normal coupling matrices.  There are two erroneous remarks in
\cite{bou} that we would like to correct in this note.

In section 4 of \cite{bou} we consider the case of a unitary mixing
matrix in the context of a model involving two families of $N$ Dirac
fermion fields each, with $N$ arbitrary, coupled to a scalar and a
massive vector field.  If the renormalized mass matrices for fermions
$\mat{M}_{j}$, $j=1,2$, are chosen in such a way that they are
hermitian in each order of perturbation theory, we can write the
relation among bare and renormalized mass matrices in the form (see
(40) in \cite{bou}),
\begin{gather}
  \mat{M}_{j\sss{0}} = \mat{U}_{mjL}^\dagger (\mat{M}_j +
  \mat{\overline{\delta} M}_j) \mat{U}_{mjR}, \label{eq:1}
\intertext{with}
  [ \mat{M}_j,\mat{\overline{\delta} M}_j ] =0, \label{eq:2}
\end{gather}
as shown in appendix A of \cite{bou}.  This is always the case in \os\ 
scheme, in which $\mat{M}_{1,2}$ are required to be real diagonal at
any perturbative order.

It is asserted in \cite{bou} that (\ref{eq:1}) and (\ref{eq:2}) hold
also in \ms\ and related schemes at one loop if the tree-level flavor
bases have been chosen so that $\mat{M}_{1,2}$ are hermitian at tree
level.  That statement is incorrect.  It is not difficult to show
that, given the polar decomposition $\mat{M}_j = \mat{R}_j \mat{V}_j$
with $\mat{R}_j$ hermitian and positive and $\mat{V}_j$ unitary, the
relation (\ref{eq:1}) holds with $\mat{\overline{\delta} M}_j =
\mat{\delta R}_j \mat{V}_j$ where $\mat{\delta R}_j$ is hermitian and
$[\mat{R}_j, \mat{\delta R}_j]=0$, that is,
\begin{equation}\label{eq:3}
  [ \mat{M}_j \mat{M}_j^\dagger, \mat{\overline{\delta} M}_j
  \mat{\overline{\delta} M}_j^\dagger ] = 0 = 
  [ \mat{M}_j^\dagger\mat{M}_j , \mat{\overline{\delta} M}_j^\dagger
  \mat{\overline{\delta} M}_j ],\quad j=1,2, 
\end{equation}
instead of (\ref{eq:2}).  The substitution of (\ref{eq:2}) by
(\ref{eq:3}) in the analysis of the model in \msb\ scheme does not
affect the other results given in \cite{bou} in any way.

Also in section 4 of \cite{bou} it is stated that due to gauge
invariance only a wave-function renormalization counterterm is needed
in the vector field sector of the model.  This is true in \ms\ scheme
but, of course, other, finite counterterms are needed in \os\ scheme.
The vector boson mass and gauge parameter renormalization, however,
are not needed in the analysis presented in that section.